# Toolbox for analyzing finite two-state trajectories

O. Flomenbom, and R. J. Silbey

***Chemistry Department, MIT, 77 Massachusetts Ave., Cambridge, Massachusetts 02139, USA***

**ABSTRACT** In many experiments, the aim is to deduce an underlying multi-substate *on-off* kinetic scheme (KS) from the statistical properties of a two-state trajectory. However, a two-state trajectory that is generated from an *on-off* KS contains only partial information about the KS, and so, in many cases, more than one KS can be associated with the data. We recently showed that the optimal way to solve this problem is to use canonical forms of reduced dimensions (RD). RD forms are *on-off* networks with connections only between substates of different states, where the connections can have non-exponential waiting time probability density functions (WT-PDFs). In theory, only a single RD form can be associated with the data. To utilize RD forms in the analysis of the data, a RD form should be associated with the data. Here, we give a toolbox for building a RD form from a *finite* time, noiseless, two-state trajectory. The methods in the toolbox are based on known statistical methods in data analysis, combined with statistical methods and numerical algorithms designed specifically for the current problem. Our toolbox is self-contained - it builds a mechanism based *only* on the information it extracts from the data, and its implementation is fast (analyzing a $10^6$ cycle trajectory from a thirty-parameter mechanism takes a couple of hours on a PC with a 2.66 GHz processor). The toolbox is automated and is freely available for academic research upon electronic request.

## I. INTRODUCTION

Finding an underlying mechanism from a binary time-series (Fig. 1) is a problem that appears in many fields in physical chemistry and biophysics (1-33), ranging from studies on the photo-physical properties of nano-crystals (21-27), studies on the structural changes and the activity of single biopolymers and small organic molecules (7-20, 28-31), to numerical studies of complex systems, e.g. protein folding and reactions (32-33). In many cases, the underlying mechanism is (assumed to be) a multi-substate *on-off* Markovian kinetic scheme (KS) (38-40); examples for KS are shown in Fig. 2: the KSs are 2A, 2C, 2E. (In this paper, we call the KS by the figure it is shown in.) That is, we assume that the mechanism is a network with a frozen connectivity. The network can be of any size with any kind of wiring (e.g. irreversible transitions are also allowed). The assumption that the true underlying mechanism that generates the data is a KS is the only assumption made here. It is the assumption that is usually made in the interpretation of binary trajectories. For example, in single enzyme kinetics experiments (19, 87), the network represents the space of the conformations of the enzyme and the reaction pathways. In translocation of molecules through channels (3, 50-57), the network represents the propagating molecule through the channel and the effect of the environment on the translocation.

The connection between the underlying mechanism and the experiment is as follows: each substate in the network has a unique observable value that belongs to one of the two states (defined as the *on* state and the *off* state). The observed two-state trajectory is generated by the random walk in the *on-off* KS, in which only transitions between substates of different states are explicitly observed. (Here, we also call the binary time-series, a two-state trajectory, or just a trajectory. The periods in the trajectory are also called events or waiting times.)

Simulating the data from a KS is straightforward; however, deducing the underlying multi-substate KS that generated the data is, almost always, not possible, even when 'analyzing' an infinitely long trajectory, because the projection of the multidimensional KS onto the two-dimensional data leads to a loss of information about the structure of the underlying KS, so two, or several, KSs can lead to identical trajectories in a statistical sense (41-47). The best strategy to deal with this situation is to use canonical forms (41-47). A given KS is mapped into a single canonical form, but many KSs can be mapped to the same canonical form. (This correspondence of many KSs to a single canonical form is a quantification of the concept stating that a single two-state trajectory does *not* (almost always) contain *all* the information about the *on-off* KS that generated it.) The canonical form should be built from the data.

Recently, we have found a map of KSs into new canonical forms, which we called reduced dimensions (RD) forms (47). A RD form is an *on-off* network with connections only between substates of different states. A RD form has the simplest topology that can generate the data, namely, it has the minimal number of substates that can generate the data. The numbers of *on* and *off* substates in the RD form are determined by the ranks of the two-dimensional (2D) waiting time probability density functions (WT-PDFs) of successive events, $\phi_{x,y}(t_1,t_2)$, *x*,*y*= *on*,*off* (47, 103). The price for having the simplest possible topology is complex of the WT-PDFs for the connections in the RD form, $\varphi_{x,ij}(t)$ for connecting substates $j_x \to i_y$: these WT-PDFs are almost always multi-exponentials in contrast to single exponential WT-PDF for connecting substates in KSs. RD forms have many advantages over the previously suggested canonical forms (for *on-off* KSs) (41-42, 46). Two of these advantages are (1) the ability of RD forms to represent any KS, i.e. also KSs with symmetry* and irreversible transitions can be mapped into RD forms, and (2) the use of RD forms as an efficient tool, more efficient than other existing tools, in discriminating between KSs by analyzing the topology of the KSs. For example, the RD forms of KSs 2A, 2C and 2E, shown in Figs. 2B, 2D and 2F, respectively, are all distinct, so their conjugate KSs can be discriminated by the analysis of the data. An elaborated discussion regarding the theoretical mapping of KSs into RD forms, and the utilities of RD forms are given in Refs. (47, 103). This paper focuses on the construction of the RD form from finite data, and is complementary to the theory involving RD forms and *on-off* KSs presented in (47, 103).

There is a vast literature dealing with the analysis (34-82) and the modeling (83-102) of experimental and theoretical two-state trajectories. However, none of the existing methods are designed to build a canonical form directly from a two-state trajectory. So, these works don't deal with several difficulties that arise when analyzing the data with canonical forms. In fact, many authors ignore main problems in analyzing time-series by assuming that substantial amount of information is already given to begin with. In this paper, we give a toolbox for the *direct* construction of the RD form from a *finite*, noiseless, two-state trajectory. To do so, we introduce several new techniques in the analysis of the data, which are combined with known statistical methods in data analysis and numerical algorithms (48-82). Among the important quantities that are extracted from the data and are unique to this toolbox are the matrices $\sigma_{x,y}$ that appear in the expansion of the $\phi_{x,y}(t_1,t_2)$ and the ranks of these matrices (which are estimated independently). An important concept in the basis of our toolbox finds from the data the *optimal* initial conditions for the mechanism-dependent optimization in the last step of the analysis. Accurate initiations of optimization routines are crucial to the success of the analysis, because the mechanism-dependent optimization always finds the nearest minimum in a very rugged parameter space. Our toolbox succeeded much better than existing methods in a set of cases we tested, and we attribute this mainly to our way of initiating the optimization subroutines. The toolbox presented in this paper is automated (Matlab codes). The toolbox is freely available for academic research upon electronic request.

---

* Symmetry means, for example, that the spectrum of a WT-PDF for single periods (*on* or *off*) is degenerate.

We emphasize that although our toolbox analyzes *noiseless* trajectories, noise filtration techniques can be applied on the raw data *before* using our toolbox. Filtering the noise in the data is a challenge specific to the way the experiment is done (see Refs. (34, 87) for examples on filtering noisy photon trajectories), but after 'cleaning' the noisy signal, the researcher faces the problem of analyzing a noiseless trajectory. This is when our toolbox is useful.

We also note that after extracting the RD form from the data, one can associate a bunch of KSs with the found RD form by using the theory of RD forms and *on-off* KSs (47, 103). Such a list of KSs may be partial, but can help in understanding the studied process. This step is independent of the analysis presented here.

This paper is organized as follows: section II introduces the reduced dimensions forms. It gives a brief summary of the results of (47, 103), and is given here for readers that are not familiar with the theory of RD forms and *on-off* KSs. Section III gives the actual toolbox to extract the canonical form from the data. We demonstrate our approach by generating the data from a particular KS (KS 3) and by analyzing it with our toolbox. The toolbox reconstructed the RD form's topology and the parameter values extracted by the toolbox are within 10% (on average) of the actual parameters. We note that this is not the only case analyzed by our toolbox that gave good results; three more tests were performed, and the accuracy of the results was, at least, the same as the one reported in this paper (we present the KS with the most complex topology among the tested KSs). The toolbox is presented in three subsections; each subsection executes a different part of the toolbox. The various parts of the toolbox are connected to each other in the last step of the analysis, which compiles the intermediate results and finds the RD form. Mathematical discussion in the main text is brief, but is given in the appendices. Section IV summaries our results.

## II. REDUCED DIMENSIONS FORMS

This section introduces the canonical forms of reduced dimensions. The section gives mainly a descriptive discussion, where the mathematical theory of reduced dimension form and *on-off* KSs is given in (47, 103).

For the following discussion, it is important to express the $\phi_{x,y}(t_1,t_2)$ s as expansion of exponentials. ($\phi_{x,y}(t_1,t_2)$ is built from the data by constructing the histogram of the intersection of successive $x$ followed by $y$ events.) The most general mathematical description of the WT-PDF $\phi_{x,y}(t_1,t_2)$ that is constructed from a two-state trajectory generated by a KS is given by (39-47):

$$\phi_{x,y}(t_1,t_2) = \sum_{i=1}^{L_x} \sum_{j=1}^{L_y} \sigma_{x,y,ij} e^{-\lambda_{x,i}t_1 - \lambda_{y,j}t_2} . \tag{1}$$

Here, we use the matrix of amplitudes, $\sigma_{x,y}$, the sets of rates, $\{\lambda_x\}$ and $\{\lambda_y\}$, and the expansion lengths, $L_x$, and $L_y$. From Eq. (1), we can construct almost any quantity of interest. For example, integrating over $t_2$ leads to,

$$\phi_x(t) = \sum_{i=1}^{L_x} c_{x,i} e^{-\lambda_{x,i}t} , \tag{2}$$

with, $c_{x,i} = \sum_{j=1}^{L_y} \sigma_{x,y,ij} / \lambda_{y,j}$. Here, $\phi_x(t)$ is the WT-PDF of the $x$ durations in the data. Note that although the above amplitudes and rates can be expressed as a function of the transition rates of the underlying KS (39-47), for the discussion below, and the analysis presented in the next section, these relationships are not useful. The reason is that RD forms are built directly from the matrices $\sigma_{x,y}$, and the rates, $\{\lambda_x\}$ and $\{\lambda_y\}$.

**Description of RD forms and some examples** RD forms are *on-off* networks with connections only between substates of different states (2B, 2D, and 2F). The topology of the RD form, i.e. the number of substates in the network, is the *simplest* topology that can reproduce the data. The trade-off is that the WT-PDFs for the connections in the RD form are (usually) multi-exponentials, in contrast to a Markovian underlying KS that has only single exponential WT-PDFs for the connections. The topology of a RD form is determined by the ranks, $R_{x,y}$ s, of the corresponding $\phi_{x,y}(t_1,t_2)$ s. For discrete time, $\phi_{x,y}(t_1,t_2)$ is a matrix with a rank $R_{x,y}$. $R_{x,y}$ is, in fact, the rank of $\sigma_{x,y}$. For non-symmetric KSs, $R_{x,y}$ for $x \neq y$ is the number of substates in state $y$ in the RD form. The WT-PDFs for the connections in the RD form, denoted by $\varphi_{x,ji}(t)$ for connecting substates $i_x \to j_y$, are determined by the mapping procedure of a KS into a RD form (47). Here, we note that $\varphi_{x,ji}(t)$ is a weighted sum of exponentials with rates, $\{\lambda_x\}$, and as many as $L_x$ components,

$$\varphi_{x,ji}(t) = \sum_{k=1}^{L_x} \alpha_{x,jki} e^{-\lambda_{x,k} t} .$$

It is straightforward to get the amplitudes, $\alpha_{x,jki}$s, and the rates numerically, by mapping a KS into a RD form (47). (The mapping of *on-off* KSs into RD forms is based on the path representation of the $\phi_{x,y}(t_1,t_2)$s (47).) Estimating the amplitudes and rates from the data is a much harder task. Our toolbox presented in Section III is designed to find the RD form from finite binary data. Lastly, we note that RD forms are canonical forms in the sense that only one RD form can be constructed from an infinitely long two-state trajectory, and this RD form contains all the information in the two-state trajectory. RD forms are canonical forms of KSs because a given KS is mapped to a unique RD form.

The simplest topology for a RD form (2B) has one substate in each of the states, namely, $R_{x,y} = 1$ ($x, y = on, off$ ). Therefore, $\varphi_{x,11}(t) = \phi_x(t)$. For a two by two RD form (2D), e.g. when $R_{x,y} = 2$ , $x, y = on, off$ , there are as many as 4 different $\varphi_{x,ji}(t)$s for each value of *x*. In general, for a RD form with $L_{RD,x}$ substates in state *x*, there are as many as $2L_{RD,on}L_{RD,off}$ different WT-PDFs for the connections in the RD form.

The basic utilities of RD forms include: (*a*) A RD form has the simplest topology that can reproduce the data. (*b*) The topology of the RD form is obtained from the data without fitting. (*c*) RD forms can represent KSs with symmetry and irreversible transitions because these canonical forms are built from all four $R_{x,y}$ s. (*d*) RD forms constitute a convenient and powerful tool, a much more powerful tool than other methods, for discriminating among *on-off* KSs.

For an elaborated discussion on the mapping of KSs into RD forms and the other utilities of RD forms, see (47, 103).

## III – TOOLBOX FOR BUILDING THE RD FORM FROM FINITE DATA

This section outlines our toolbox for a direct construction of the RD form from a *finite*, noiseless, trajectory. The concepts behind the toolbox and the methods of analysis are discussed in the main text .Complementary mathematical details and statistical significance tests are given in appendices A-C. Our toolbox executes a four-step algorithm:

(**1**) Estimation of the spectrum and amplitudes of the $\phi_x(t)$ s using fitting procedures.

(**2**) Determination of the number of substates in the RD form from the estimated ranks of the $\phi_{x,y}(t_1,t_2)$ s.

(**3**) Estimation of the matrices $\sigma_{x,y}$ s in the exponential expansion of the $\phi_{x,y}(t_1,t_2)$ s.

(**4**) Determination of the RD form. The starting point is a low resolution RD form, which is built from the information collected in the first stages of the analysis. This step determines the pre-exponential coefficients of the WT-PDFs for the connections in the RD form by an optimization procedure. The optimization subroutine uses the matrices $\sigma_{x,y}$ s, and the rates $\{\lambda_{on}\}$ and $\{\lambda_{off}\}$.

Our toolbox is based on a new approach for analyzing the trajectory: it builds the matrices $\sigma_{x,y}$, estimates the ranks of $\phi_{x,y}(t_1,t_2)$, and phrases the first mechanism-dependent optimization step as a root-search problem. To the best of our knowledge, none of these techniques were used before in the analysis of two-state trajectories, but, here, were proven crucial for the success of the analysis, i.e. constructing a reliable canonical form of reduced dimensions.

The toolbox presented in this section, together with its accompanied appendices, give efficient methods for carrying out the above algorithmic steps. Each step in the above algorithm solves a problem that is challenge by its own, but, importantly, the methods and intermediate results are combined in self-contained codes to give the final output of an optimal RD form that can be associated with the input two-state trajectory.

To present the course of the analysis, we apply the toolbox on the data generated by KS 3. This KS has two loops and two irreversible transitions. We also impose symmetry in the KS: the splitting probabilities obey $p_{2_{on}2_{off}} = p_{2_{on}1_{off}}$ and $p_{1_{on}2_{off}} = p_{1_{on}1_{off}}$, where the splitting probability $p_{ij}$ is defined by, $p_{ij} = k_{ij} / \sum_i k_{ij}$, and $k_{ij}$ connects substates $j$ to $i$. The corresponding canonical form's topology of KS 3 is RD form 2D. The particular RD from of KS 3 has four bi-exponential and four single exponential WT-PDFs for the connections: $\varphi_{on,11}(t)$ and $\varphi_{on,21}(t)$ are single-peaked WT-PDFs with two components and are proportional to each other, $\varphi_{on,12}(t)$ and $\varphi_{on,22}(t)$ are both single-component WT-PDFs, but are not proportional. For the *off* state, $\varphi_{off,11}(t)$ and $\varphi_{off,21}(t)$ are proportional WT-PDFs with two components, and $\varphi_{off,12}(t)$ and $\varphi_{off,22}(t)$ are proportional WT-PDFs with one component. (Table 4 gives the numerical values for the expansion parameters for all the $\varphi_{x,ij}(t)$s.) KS 3 and its corresponding RD form are complex enough to serve as a good test for our toolbox. We use a $10^6$ event trajectory, which is a typical size trajectory from ion-channel recordings.

**III.1 The WT-PDFs of the single events** The first stage in the analysis of a two-state trajectory constructs the WT-PDF of state *x* (=*on*, *off*), $\phi_x(t)$, by building the histogram from the *x* periods of the trajectory. $\phi_x(t)$ gives basic information about the dynamics and the KS, deduced by finding its functional form. For a Markovian KS, $\phi_x(t)$ is a sum of exponentials, $\phi_x(t)=\sum_{i=1}^{L_x} c_{x,i}e^{-\lambda_{x,i}t}$, so the functional form of $\phi_x(t)$ is completely determined by specifying the $c_{x,i}s$ and the $\lambda_{x,i}s$. The spectrum $\{\lambda_x\}$ of $\phi_x(t)$ is the same spectrum of the WT-PDFs for the connections in the RD form, and the number of exponentials in $\phi_x(t)$, $L_x$, estimates the number of substates in state *x* in the underlying KS.

There is a vast literature on estimating the coefficients and rates in an exponential expansion of an experimental curve (58-68). We have developed a method based on the Padé approximation approach (60-64, 68), but our method also directly maximizes a mechanism-free likelihood function. (In this paper, we use the term 'mechanism-free optimization' for cases in which the optimization is not related to a KS or to a RD form.) The method is robust numerically, and estimates not just the $c_{x,i}s$ and the $\lambda_{x,i}s$ but also the optimal $L_x$. The details of the subroutine are given in appendix A.

Figure 4 shows the $\phi_x(t)$s obtained from a trajectory of $10^6$ *on-off* events generated by KS 3. Also shown are the analytical curves and the curves found by our subroutine. The subroutine found the correct number of components in both fits. Table 1 gives the numerical values of the parameters found by the subroutine and the corresponding analytical values. Although $\phi_{on}(t)$ has a fairly complicated shape, both the fitting amplitudes and fitting rates are within 5% of the analytical values. The fit for $\phi_{off}(t)$ yields good results also, but here the maximal error (occurring in the smallest rate and its conjugated amplitude) is about 20%. (The error can be reduced by enlarging the data set.)

**III.2 Degree of correlations in the 2D histograms and the matrices $\sigma_{x,y}$** The next two algorithmic steps in the toolbox estimate the degree of correlations between successive events and the matrices $\sigma_{x,y}$ s. This information, together with the rates $\{\lambda_{on}\}$ and $\{\lambda_{off}\}$ found in the first step of the analysis, enables building a low-resolution RD form, and to initialize properly the mechanism–dependent optimization in the last step of the analysis. Constructing the matrices $\sigma_{x,y}$ and the $R_{x,y}$ is unique for our toolbox.

The degree of correlations between *x* event followed by *y* event is the rank $R_{x,y}$ of $\phi_{x,y}(t_1,t_2)$. When, $R_{x,y}=1$, the data is not correlated, and for $R_{x,y}>1$ there are correlations between events in the data of order $R_{x,y}$. The order of the correlations is the minimal number of 2D-functions needed to expand, $\phi_{x,y}(t_1,t_2)$. Because this expansion is some kind of a linear decomposition of $\phi_{x,y}(t_1,t_2)$, the rank $R_{x,y}$ can be obtained by any kind of linear decomposition; so the number of non-zero eigenvalues in a standard linear decomposition of the matrix $\phi_{x,y}(t_1,t_2)$ is $R_{x,y}$. (Note that these eigenvalues are not the $\{\lambda_x\}$.) When dealing with finite data, the rank $R_{x,y}$ is not easily determined; one cannot simply analyze the spectrum of the experimental $\phi_{x,y}(t_1,t_2)$, because this spectrum contains more non-zero entries than the rank of the noiseless $\phi_{x,y}(t_1,t_2)$. As a result, it is useful to begin with an analysis that finds whether events are correlated or not, because this information is easily obtained from the finite data with a high degree of certainty. Subsection III.2.1 gives six different methods for determining whether correlations between events exist in the data. These methods are fast (implementation takes a couple of seconds for a $10^6$ event trajectory from a fifteen parameters KS), but depending on the data's properties (e.g. length, complexity, pattern of correlations), different approaches are more suitable than others for a given data.

The analysis in subsection III.2.1 also obtains the matrices $\sigma_{x,y}$ s. These matrices are used in the mechanism-dependent optimization procedure in the last step of the construction of the RD form from the data. The construction of the matrices $\sigma_{x,y}$ is unique for our toolbox.

Subsection III.2.2 gives the method that estimates the $R_{x,y}$ values, which is only applied on non-renewal data, namely, when the trajectory shows correlations between successive events (otherwise, all ranks equal unity, i.e. $R_{x,y} = 1$ for all four combinations of *x* and *y*).

**III.2.1 Determining the existence or lack of correlations in the data and the matrices $\sigma_{x,y}$ s** In this subsection, six ways to detect correlations between events in the trajectory are introduced. Different techniques are useful in analyzing different data types, but none of these demand a direct construction of the 2D histograms. Two of these techniques construct special single-argument WT-PDFs from successive events. It is shown that these special WT-PDFs contain all the information necessary to build the matrices $\sigma_{x,y}$ s. The six methods for detecting correlation in the data are described and applied on numerical trajectory generated by KS 3. Mathematical discussion and statistical significance tests needed for implementing the methods in a routine are given in appendix B.

(**a**) A simple way to detect correlations between events in the data analyzes the trajectory of waiting times plotted vertically as a function of the occurrence index (43, 79, 87), see Fig. 5A. This trajectory is called the ordered waiting times trajectory. Ways to analyze the ordered waiting times trajectory are given below.

(**a.1**) For some cases, correlations between successive events can be easily detected *visually* in the ordered waiting times trajectory (43, 79, 87). This happens when waiting times of similar duration are grouped, but different groups have very different average durations.

(**a.2**) The correlation function of the ordered waiting times trajectory can be calculated (17, 40, 79, 87), $< t_x t_y(m) >= \frac{1}{N_{tra}-m}\sum_{i=1}^{N_{tra}-m} t_{x,i} t_{y,i+m}$ , where $t_{z,j}$ is the $j^{th}$ event in state *z* and $N_{tra}$ is the number of cycles in the trajectory. The data is correlated when this correlation function is *not* the Kronecker delta, $\delta_{0m}$. However, this correlation function is not always useful, and in some cases can lead to false-negative results (79). There are (at least) two reasons for this behavior. Firstly, note that the argument of the correlation function of the ordered waiting times trajectory represents the distance, in number of events, between the two events in the 2D WT-PDFs, so its argument is an integer and its value at a given integer is the first moment of the corresponding 2D WT-PDF (40, 79). Now, because the statistics of 2D WT-PDFs of distant (i.e. not successive) events are not accurately obtained from finite trajectories, the correlation function of the ordered waiting times trajectory is noisy even for small argument values. The second reason that causes this correlation function to fail involves specific situations of data made of short correlated events combined with long uncorrelated events.

(**a.3**) Instead of looking on the correlation function of the ordered waiting time trajectory, the statistics of the 2D WT-PDF of *successive* events can be calculated. For example, the moments of $\phi_{x,y}(t_1,t_2)$, $<t_{1,x}^m t_{2,y}^n>$ ($n$, $m$ > -1), can be computed directly from the ordered waiting times trajectory and compared with the moments of the product $\phi_x(t_1)\phi_y(t_2)$, $<t_x^m><t_y^n>$ (43), which are also computed directly from the data. When $<t_{1,x}^m t_{2,y}^n> \neq <t_x^m><t_y^n>$, the data is correlated. This technique may be more robust than the correlation function in (**a.2**), because it probes only the properties of $\phi_{x,y}(t_1,t_2)$. However, it can also fail for small data sets, and for data sets made of short correlated events combined with long uncorrelated events (79).

(**a.4**) The correlation function of the *on-off* trajectory is computed directly from the data and is compared with the correlation function of a renewal process. The latter correlation function is obtained from the experimental $\phi_{on}(t)$ and $\phi_{off}(t)$ (87), or from the randomized data. (Here, by randomized data we mean a time series in which the waiting times in the original trajectory are randomly permutated, such that the obtained trajectory is a renewal process.) If there are significant differences between the two correlation functions, the data is correlated. Although this method also answers whether the data is correlated or not, it doesn't analyze the ordered waiting times trajectory, and so it is heavier computationally-wise relative to the methods (**a.1**) - (**a.3**). (The trajectory value at each time increment enters in these calculations, rather than the random durations.)

Techniques (**a.1**)-(**a.3**) were applied to the data generated from the KS 3. The ordered waiting time trajectory (Fig. 5A) shows the typical signature for correlations, here *off-off* correlations, as bunching of successive short events followed by bunching of successive long events. The event correlation function for *off-off* events supports this conclusion (Fig. 5B). The moment analysis is applied on *on-on* events, and indicates on correlations between *on-on* events (Fig. 5C). Note that visual inspection of the ordered waiting time trajectory is not as sensitive as the event correlation function and the moment analysis, but can be useful when dealing with small data sets. Additionally, it is evident from the error bars in Figs. 5B-5C that the event correlation function becomes noisier much faster (with the distance between events) relative to the moments of a 2D histogram of successive events (with the moments' order).

As a technical note we emphasize that to use the above methods in a routine (for example the moment analysis), one must define a significance level for evaluating the results. A way to define a significance level assumes Poissonian noise in the moment values with the trajectory's length, and compares the results obtained from the data with the results obtained from the randomized data. In appendix B, we give the details of the analysis used in our routine.

(**b**) A different approach for detecting correlations between events in finite data builds special single-argument WT-PDFs. These WT-PDFs represent sums (of different functions) of the successive periods. When comparing such a special WT-PDF with a similar WT-PDF that is calculated from the randomized-data, it is easy to determine whether the data is correlated or not by a visual inspection. (Only when the two different WT-PDFs are statistically equivalent is the data uncorrelated; see Fig. 6 for examples.) These special WT-PDFs are more robust than the techniques in (**a.1**)-(**a.3**) for detecting correlations between events (79) (the reasons are given below). Importantly, we have developed a method that uses the information contained in these special WT-PDFs to build the matrices $\sigma_{x,y}$ s. This method is spelled out in this subsection. We emphasize that the matrices $\sigma_{x,y}$ s are crucial to the successes of the analysis and are unique to this toolbox (namely, it is shown here for the first time how to extract these matrices from the *on-off* time-series).

(**b.1**) Consider the WT-PDF obtained from the data by building the histogram of the random waiting time that is a sum of successive waiting times, $\left\{t_i = t_{x,i} + t_{y,\delta_{x,y}+i}\right\}_{i=1}^{N_{tra}-\delta_{x,y}}$, where $t_{z,j}$ is the $j^{th}$ event in state $z$ and $N_{tra}$ is the number of cycles in the trajectory. We call this WT-PDF the binned successive WT-PDF, and is denoted by $\phi_{x+y}(t)$. $\phi_{x+y}(t)$ is built from all available data points, has a single argument, and is related only to the PDF of successive periods. These three points make $\phi_{x+y}(t)$ more robust than the methods described in (**a**) for detecting correlations in the data. $\phi_{x+y}(t)$ is defined by,

$$\phi_{x+y}(t) = \int_0^\infty \int_0^\infty \delta[t-(t_1+t_2)]\phi_{x,y}(t_1,t_2)dt_1dt_2 \ . \tag{3}$$

$\phi_{x+y}(t)$ is compared with the convolution of $\phi_x(t)$ and $\phi_y(t)$, denoted by $\phi_x(t) * \phi_y(t)$, which is obtained from the data by performing the same procedure defined above for $\phi_{x+y}(t)$ *after* randomizing the data, or by a numerical integration. The convolution $\phi_x(t) * \phi_y(t)$ is expressed analytically by replacing $\phi_{x,y}(t_1,t_2)$ in the integrand in Eq. (3) with $\phi_x(t_1)\phi_y(t_2)$. For the example discussed here, Fig. 6A shows the strength of the comparison method of PDFs relative to methods in (**a**): the two curves corresponding to $\phi_{on+off}(t)$ and $\phi_{on}(t) * \phi_{off}(t)$ are so different, and therefore it is straightforward to determine that the data is correlated (*on-off* correlations).

Going back to Eq. (3) and integrating over one variable leads to,

$$\phi_{x+y}(t) = \int_0^t \phi_{x,y}(t-\tau,\tau)d\tau \ ,$$

because $\phi_{x,y}(t_1,t_2)$ with negative arguments is zero. Thus, $\phi_{x+y}(t)$ is obtained by integrating over a straight line in the 2D plane that intersects the axes in $(0,t)$ and $(t,0)$. As a result, for small *t*, $\phi_{x+y}(t)$ probes correlations only between short periods. For large *t*, $\phi_{x+y}(t)$ probes correlations between long periods, as well as correlations between long-short periods. This makes $\phi_{x+y}(t)$ a good method for detecting correlations between short successive events in a data that also has long uncorrelated periods, a scenario that can fail the techniques presented in (**a**).

$\phi_{x+y}(t)$ can be used for detecting correlations between successive events in a trajectory generated from any mechanism. This simple test doesn't require knowing the functional form of $\phi_{x+y}(t)$. However, the actual functional form of $\phi_{x+y}(t)$ is important for finding the matrices $\sigma_{x,y}$. For trajectories from KSs, we can express $\phi_{x+y}(t)$ by inserting $\phi_{x,y}(t_1,t_2)$ in Eq. (1) into the integral representation of $\phi_{x+y}(t)$ in Eq. (3). Integration gives,

$$\phi_{x+y}(t) = \sum\nolimits_{i,j} \sigma_{x,y,ij}(e^{-\lambda_{x,i}t} - e^{-\lambda_{y,j}t})/(\lambda_{y,j} - \lambda_{x,i}) \ . \tag{4}$$

When analyzing the data, it is important to know how many different terms are contained in $\phi_{x+y}(t)$. It is easily seen that Eq. (4) expresses $\phi_{x+y}(t)$ as a weighted sum of $L_{on} + L_{off}$ exponentials by rewriting Eq. (4) in the form,

$$\phi_{x+y}(t) = \sum_{i=1}^{L_x} \sigma_{x+y,i} e^{-\lambda_{x,i} t} + \sum_{j=1}^{L_y} \sigma_{x+y,j} e^{-\lambda_{y,j} t} , \tag{5}$$

where $\sigma_{x+y,i} = \sum_{j=1}^{L_y} \sigma_{x,y,ij} / (\lambda_{y,j} - \lambda_{x,i})$ and $\sigma_{x+y,j} = \sum_{i=1}^{L_x} \sigma_{x,y,ij} / (\lambda_{x,i} - \lambda_{y,j})$. (Note that the first sum involves summation over the columns of $\sigma_{x,y}$, but the second sum is done over its rows; namely, the transpose relation exists between the two sums. We keep this distinction implicit in notation.) Using Eq. (5), we can recover the $\sigma_{x+y,i}$ s from an experimentally obtained $\phi_{x+y}(t)$. To get the amplitudes of $\phi_{x+y}(t)$, we use the exponential fit command presented in Section III.1. We automated the construction of the exponential expansion of $\phi_{x+y}(t)$ from a trajectory. Appendix B.2 discusses possible difficulties and their solutions when building the $\sigma_{x+y,i}$ s from the data. (This appendix also uses the $\sigma_{x+y}$ s for finding the matrices $\sigma_{x,y}$.) The results for the construction of the $\sigma_{x+y}$ s, $x \neq y$, from the trajectory of KS 3 are given in table 2, and the corresponding WT-PDFs are shown in Fig. 6A. In the end of the next subsection we discuss these results.

(**b.2**) Another special single-argument WT-PDF is the histogram of the random time that is the sum of the square root of successive periods, $\left\{\sqrt{t_i} = \sqrt{t_{x,i}} + \sqrt{t_{y,\delta_{x,y}+i}}\right\}_{i=1}^{N_{tra}-\delta_{x,y}}$. This function is defined by,

$$\phi_{\sqrt{x}+\sqrt{y}}(\sqrt{t}) = \int_0^\infty \int_0^\infty \delta[\sqrt{t} - (\sqrt{t_1} + \sqrt{t_2})] \phi_{x,y}(t_1, t_2) dt_1 dt_2 . \tag{6}$$

We call functions of the type of $\phi_{\sqrt{x}+\sqrt{y}}(\sqrt{t})$ the generalized-binned successive WT-PDF because it is a generalization of $\phi_{x+y}(t)$, which involves a linear sum of successive events, to any sum of non-linear functions of the successive events. $\phi_{\sqrt{x}+\sqrt{y}}(\sqrt{t})$ is compared with a function obtained by replacing $\phi_{x,y}(t_1, t_2)$ with $\phi_x(t_1)\phi_y(t_2)$ in Eq. (6), and is constructed from the data by performing the same operation defined above for $\phi_{\sqrt{x}+\sqrt{y}}(\sqrt{t})$ *after* randomizing the data, or by a numerical integration. Using the data, two generalized-binned successive WT-PDFs, obtained from the data and the randomized data for *off-on* events, are shown in Fig. 6B. The differences between the two curves are apparent, meaning that *off-on* events are correlated.

Going back to the integral representation of $\phi_{\sqrt{x}+\sqrt{y}}(\sqrt{t})$ and performing the integration over the delta function leads to,

$$\phi_{\sqrt{x}+\sqrt{y}}(\sqrt{t}) = 2\int_0^t \left(\sqrt{t}-\sqrt{\tau}\right)\phi_{x,y}\left[\left(\sqrt{t}-\sqrt{\tau}\right)^2,\tau\right]d\tau\,, \tag{7}$$

and can be written as,

$$\phi_{\sqrt{x}+\sqrt{y}}(\sqrt{t}) = 4t^{3/2}\int_0^1 z(1-z)\phi_{x,y}\left(t(1-z)^2, tz^2\right)dz\,. \tag{8}$$

Namely, $\phi_{\sqrt{x}+\sqrt{y}}(\sqrt{t})$ is obtained by integrating over curved lines, $y(x)=(t^{1/2}-x^{1/2})^2$, in the 2D plane. This enables detecting correlations between successive waiting times along the axes more efficiently than $\phi_{x+y}(t)$. Note that the generalized binned successive WT-PDF in Eqs. (6), (7), (8), is a particular example of a family of WT-PDFs obtained by integrating the 2D WT-PDFs with the delta function, $\delta(t^\mu - t_1^\mu - t_2^\mu)$. Simple analysis shows that all the important information can be extracted from the $\phi_{\sqrt{x}+\sqrt{y}}(\sqrt{t})$ s.

The last part of this subsection presents the method that uses $\phi_{\sqrt{x}+\sqrt{y}}(\sqrt{t})$ s for estimating the matrices $\sigma_{x,y}$. Firstly, we express $\phi_{\sqrt{x}+\sqrt{y}}(\sqrt{t})$ for an underlying KS as,

$$\phi_{\sqrt{x}+\sqrt{y}}(\sqrt{t}) = 4t^{3/2}\sum\nolimits_{i,j}\sigma_{x,y,ij}e^{-\lambda_{x,i}\lambda_{y,j}t/(\lambda_{x,i}+\lambda_{y,j})}I_{ij}(t)\,, \tag{9}$$

where, $I_{ij}(t)=\int_0^1 z(1-z)e^{-r_{ij}t(z-l_{ij})^2}dz$, with, $r_{ij}=\lambda_{x,i}+\lambda_{y,j}$ and $l_{ij}=\lambda_{y,j}/r_{ij}$. The method for estimating the sigma matrices uses fitting procedures and for this, it is more convenient to work with a related quantity, $\phi_{\left(\sqrt{x}+\sqrt{y}\right)^2}(t)/(2t)$, rather than with Eq.(9),

$$\phi_{\left(\sqrt{x}+\sqrt{y}\right)^2}(t)\Big/2t = \sum\nolimits_{i=1}^{L_x}\sum\nolimits_{j=1}^{L_y}\sigma_{x,y,ij}e^{-\lambda_{x,i}\lambda_{y,j}t/(\lambda_{x,i}+\lambda_{y,j})}I_{ij}(t)\,, \tag{10}$$

because it has a larger density of points than $\phi_{\sqrt{x}+\sqrt{y}}(\sqrt{t})$. $\phi_{\left(\sqrt{x}+\sqrt{y}\right)^2}(t)$ is built from random times obtained by squaring the sum of the square root of successive events, i.e. $\left\{t_i=\left(\sqrt{t_{x,i}}+\sqrt{t_{y,\delta_{x,y}+i}}\right)^2\right\}_{i=1}^{N_{tra}-\delta_{x,y}}$. Note that $\phi_{x+y}(t)$ contains up to $L_{on}+L_{off}$ detectable terms, but $\phi_{\left(\sqrt{x}+\sqrt{y}\right)^2}(t)/(2t)$, and the related quantity, $\phi_{\sqrt{x}+\sqrt{y}}(\sqrt{t})/(4t^{3/2})$,

contains up to $L_{on}L_{off}$ detectable terms, so the latter PDF gives, in principle, more information on the $\sigma_{x,y,ij}$ s than the former.

To find the $\sigma_{x,y,ij}$ s, we need to fit $\phi_{\left(\sqrt{x}+\sqrt{y}\right)^2}(t)/(2t)$ and use Eq. (10). The exponential fit subroutine of subsection III.1 cannot be used to recover the $\sigma_{x,y,ij}$ s from Eq. (10) because the $I_{ij}(t)$ s depend on *t*. However, by an asymptotic short time expansion to the first order of the exponent in $I_{ij}(t)$, Eq. (10) can be approximated as,

$$\phi_{\left(\sqrt{x}+\sqrt{y}\right)^2}(t) \Big/ {t/3} \approx \sum_{i=1}^{L_x} \sum_{j=1}^{L_y} \sigma_{x,y,ij} e^{-0.3t(\lambda_{x,i}+\lambda_{y,j})},$$

valid when $t(\lambda_{x,i}+\lambda_{y,j})<1$ for every *i* and *j*. In the subroutine that estimates the matrix $\sigma_{x,y}$, we first fit the asymptotic expansion of $\phi_{\left(\sqrt{x}+\sqrt{y}\right)^2}(t)/(2t)$ to a sum of exponentials. Such a fit provides only a partial estimation of the $\sigma_{x,y,ij}$ s. To proceed, we designed a numerical algorithm that builds a matrix from the $c_x$ s, the $\sigma_{x+y}$ s, and the fit amplitudes of $\phi_{\left(\sqrt{x}+\sqrt{y}\right)^2}(t)/(2t)$, and uses a singular value decomposition to estimate all the elements in the matrix $\sigma_{x,y}$. This estimation is then used as an initial condition in a mechanism-free maximum likelihood procedure with constraints to get the final estimation for the $\sigma_{x,y,ij}$ s. The object likelihood function is constructed from Eq. (1). The constraints are built from the coefficients $c_x$ s and the $\sigma_{x+y}$ s, and by demanding the positivity of the WT-PDF. Appendix B.2 gives additional details, and discusses difficulties and their solutions in estimating the $\sigma_{x,y,ij}$ s from the data.

$\phi_{on+off}(t)$ and $\phi_{(\sqrt{off}+\sqrt{on})^2}(t)$ are constructed from the data generated by KS 3, and are plotted in Figs. 6A-6B, respectively. Each WT-PDF is compared with the result of the randomized data. In both cases, the fit curves for the randomized data (upper curves) are obtained from the parameters of the $\phi_x(t)$ s. There are two fit curves for each of the actual WT-PDFs (lower curves): in both cases, one fit curve is found by the exponential fit subroutine, where the second fit curve is obtained by constructing the $\sigma_{on+off}$ s for $\phi_{on+off}(t)$ and the matrix $\sigma_{off,on}$ for $\phi_{(\sqrt{off}+\sqrt{on})^2}(t)$. In both cases, the fit curves coincide with each other and with the data. Table 2 gives the analytical and fit parameters for the $\phi_{on+off}(t)$ and $\phi_{off+on}(t)$. The mean error, for both $\sigma_{x+y}$ s is about twenty percent, excluding the amplitude with the largest error. For both $\sigma_{x+y}$s, the largest error occurs in the smallest amplitude and is about a factor of 2.5. Table 3 gives the analytical and fit amplitudes for $\sigma_{x,y}$s. The coincidence is satisfactory: the average error is forty percent for $\sigma_{on,off}$ and twenty six percent for $\sigma_{off,on}$. We note that there are solutions for the $\sigma_{x,y}$s that can reproduce $\phi_{x+y}(t)$ but not $\phi_{(\sqrt{x}+\sqrt{y})^2}(t)$. (The subroutine for obtaining the $\sigma_{x,y}$s produces an ensemble of possible solutions and chooses the best among them; see appendix B for details).

Thus, the role of the $\phi_{x+y}(t)$s and $\phi_{\sqrt{x}+\sqrt{y}}(\sqrt{t})$s in our routine is two-fold: not just that these WT-PDFs indicate, very efficiently, whether the data is correlated or not, these WT-PDFs are also used to construct the $\sigma_{x,y,ij}$s. To the best of our knowledge, the way to construct the matrices $\sigma_{x,y}$s from a two-state trajectory is new. These matrices are essential for a reliable estimation of the coefficients of the WT-PDFs for the connections in the RD form.

**III.2.2 Finding the ranks of the 2D histograms** The methods described in **III.2.1** can detect correlations between events in the data. When correlations between events are found, the ranks of the 2D histograms should be estimated (no correlations between events means that all the ranks equal unity). A method that estimates the exact degree of the correlation between successive events is presented in this subsection, and it is based on the analysis given in Refs. (47, 69-70) (the problem of rank estimation of a 2D histogram is closely related to problems that arise in the field of signal reconstruction from noisy data, e.g. voice and image reconstruction, and some of the above references are related to these problems.) We note that, mathematically, the rank of $\sigma_{x,y}$ is the rank of the corresponding $\phi_{x,y}(t_1,t_2)$. However, statistical errors in $\sigma_{x,y}$ cannot be corrected, and can lead to wrong conclusions. The best way to deduce the topology of the RD form is by estimating the rank of the 2D-histograms directly from the data, and a method of doing this is presented in this subsection.

We start by constructing cumulative 2D histograms. A cumulative PDF, $\phi_{x,y}(t_1,t_2;C_1)$, of a PDF $\phi_{x,y}(t_1,t_2)$ [$\equiv \phi_{x,y}(t_1,t_2;C_0)$], is defined by,

$$\phi_{x,y}(T_1,T_2;C_1) = \int_0^{T_1}\int_0^{T_2} \phi_{x,y}(t_1,t_2)dt_1dt_2 .$$

The generalization to higher order cumulative PDFs naturally follows,

$$\phi_{x,y}(T_1,T_2;C_n) = \int_0^{T_1}\int_0^{T_2}\phi_{x,y}(t_1,t_2;C_{n-1})dt_1dt_2 \ .$$

A cumulative two-dimensional PDF reduces the noise in the *original* WT-PDF, but also preserves the rank of the original PDF. (This can be seen by using the path representation of Eq. (1) (47).) For each 2D WT-PDF and its first three cumulative PDFs, we obtain the spectrum of singular values and plot the *ratio* of successive singular values as a function of the order of the large singular value in the ratio. This plot should show large values for signal ratios and a constant value of about a unity for noise ratios. Also, noise singular values must be small enough (69). This second demand forces us to work with the second order cumulative WT-PDF, rather than the first order one. The bin size used to construct the two-dimensional PDFs is found by demanding that the randomized-data two-dimensional PDFs are rank one, a preliminary procedure that was found crucial for the success of the rank estimation method.

To define a threshold that separates signal from noise ratios, we rely on small perturbations around the analytical 2D WT-PDF, and estimate the threshold value, for our problem, to lie in the range 5-7 with probability 0.99. This threshold is obtained from the bounds for the largest noise singular value in the perturbed matrix, given small enough noise level (69). (The upper bound on the largest noise singular value equals the size of the matrix times the standard deviation in the noise level, and must be smaller than the smallest singular value in the unperturbed matrix.) This estimation for the threshold yields accurate results when working with the second order cumulative 2D WT-PDF. In appendix B.3, additional technical details of the analysis used in our routine are given, with a discussion regarding the above numerical values.

The rank-estimation method is applied on the data generated by KS 3. Figures 7A-7B show the singular ratios of $\phi_{off,on}(t_1,t_2)$ and $\phi_{on,off}(t_1,t_2)$, respectively. In both cases, a large drop at the third singular ratio is observed when working with the second order cumulative WT-PDF singular. This indicates a rank two matrix. (Similar results were obtained for $\phi_{on,on}(t_1,t_2)$ and $\phi_{off,off}(t_1,t_2)$.)

We note that variation in the rank values around the found values can be considered in cases where this method gives results that are not 'sharp' enough. If this leads also to a change in the topology of the RD form, the analysis of section III.3 can be performed for each of the possible RD forms. We choose among these RD forms using information criteria.

The translation of the estimated $R_{x,y}$ s into the number of substates in the RD form, usually relies on the equality, $L_{RD,y} = \max\{R_{x,y}; R_{y,y}\}$ $(x \neq y)$, unless a particular same event rank is the largest among all ranks, and in such a case, this rank determines the number of substates in both states in the RD form. A complete table that translates any combination of relative ranks values into RD form's topology is given in (103).

**III.3 The WT-PDFs for the connections in the RD form** The final stage in the direct construction of the RD form from the data uses the information found in III.1 and III.2 in an algorithm that estimates the amplitudes of the $\varphi_{x,ji}(t)$ s. For a Markovian underlying KS, $\varphi_{x,ji}(t)$ is given by,

$$\varphi_{x,ji}(t) = \sum_{k=1}^{L_x} \alpha_{x,jki} e^{-\lambda_{x,k} t} \quad ; \quad i = 1,\dots,L_{RD,x}, \quad j = 1,\dots,L_{RD,y}. \tag{11}$$

At this stage, $L_{RD,on}$ and $L_{RD,off}$ are known from the analysis in III.2. The $\lambda_{x,i}$ s and $L_x$ s are known from the analysis in III.1. Equation (11) introduces $(L_{on} + L_{off})L_{RD,on}L_{RD,off}$ coefficients $\alpha_{z,jki}$ s, *z*=*on*, *off*. We denote the parameter space by $\mathbf{\Theta}$, $\mathbf{\Theta} = \{\alpha_{on}; \alpha_{off}\}$. The RD form is determined once $\mathbf{\Theta}$ is determined.

We estimate the $\alpha_{x,jki}$ s in a two-step subroutine. The full details are given in appendix C, and here we briefly sketch these two steps.

(***a***) We produce a set of initial coefficients, denoted by, $\{\beta_{on}; \beta_{off}\}$. The set $\{\beta_{on}; \beta_{off}\}$ is found by iterative calculations with constraints and a random updating role. The equations for updating the $\{\beta_{on}; \beta_{off}\}$ [Eqs. (C7)-(C9)] are derived by demanding that the experimentally found $c_{x,i}$ s and $\sigma_{x,y,ji}$ s are recovered from the RD form. The constraints in the iterations demand that all the WT-PDFs are non-negative, $\varphi_{x,ji}(t) \geq 0$, for every relevant value of *t*, and every *i*, *j*, and *x*. For the current problem, the non-negativity condition on a given $\varphi_{x,ji}(t)$ immediately means that $\varphi_{x,ji}(t)$ can be normalized, $\sum_j \int_0^\infty \varphi_{x,ji}(t)dt = 1$, such that $0 \leq \int_0^\infty \varphi_{x,ji}(t)dt \leq 1$. The iterative subroutine runs for $\left((L_{on} + L_{off})L_{RD,on}L_{RD,off}\right)^2$ rounds, and the best result is sorted.

(***b***) The result from the first stage (***a***) is subject to an optimization subroutine that minimizes the differences between the data-obtained $c_{x,i}$ s and $\sigma_{x,y,ji}$ s and theirs RD form values, defined, respectively, as, $\varepsilon_{x,i}$ and $\varepsilon_{x,y,ji}$. This optimization is phrased as a root-search problem. The analytical derivatives of the target function are used. The constraints in this optimization subroutine are similar to those of the iterative scheme, but are phrased as demands on the moments of the $\varphi_{x,ji}(t)$ s: $\int_0^\infty \varphi_{x,ij}(t)t^m dt \geq 0$ with $\sum_i \int_0^\infty \varphi_{x,ij}(t)dt = 1$ for every *j*, and *x*. (Usually we take two values for *m*, *m=0,* and *m=3*.)

The above two-step subroutine is performed many times, because the optimization always finds the nearest minimum, in a very rugged parameter space. We note that this two-step subroutine doesn't use the data directly, but only the information derived from it, stored in the $c_{x,i}$ s and $\lambda_{x,i}$ s and the $\sigma_{x,y,ij}$ s. However, a subsequent optimization can minimize minus log-likelihood using the data. The target function is a sum of the four $\phi_{x,y}(t_1,t_2)$ s built from the RD form, and is initialized by the set $\{\alpha_{on};\alpha_{off}\}$. Such a second optimization may improve the results when the $\sigma_{x,y,ij}$ s are relatively noisy (as can be deduced by an error analysis using the $\phi_{(\sqrt{x}+\sqrt{y})^2}(t)$ s).

Table 4 gives the final results for the $\{\alpha_{on};\alpha_{off}\}$ obtained from the routine, and compares it to the analytical results. The values for the experimental $\{\alpha_{on};\alpha_{off}\}$ are within seven percent for $\{\alpha_{on}\}$, and seventeen percent for $\{\alpha_{off}\}$, of their analytical values, excluding zeros. Zeros are harder to detect. Figure 8 compares the obtained $\phi_{(\sqrt{x}+\sqrt{y})^2}(t)$ from the model to the analytical results. The fits are satisfactory. In fact, all the fit curves coincide with the analytical curves. This remarkable coincidence means that all the matrices $\sigma_{x,y}$ s are well approximated by the RD form found by our toolbox. Numerically, the average error between entries is $10^{-8}$. The fact that the experimental set $\{\alpha_{on};\alpha_{off}\}$ does not reproduce some of zeros in the analytical set show up as negative higher moments (larger than three) for the $\varphi_{on,j1}(t)$ *j=1, 2.* (It also shows up in the value of the average error between entries ($10^{-8}$), meaning local minimum.) This result means that the RD form can indeed be uniquely constructed from the data, and this is a matter of finding the correct 'configuration' in the parameter space. Thus, the result can be improved by searching the parameter-space for longer times (the results presented here were obtained after a six hours run on a PC with ~2.66 GHz processor and 1 GB RAM), and also more efficiently. For example, one can apply a Monte Carlo procedure with the set $\{\beta_{on};\beta_{off}\}$ as an initial configuration, and look for better results with respect to the sum of the differences, $\varepsilon_{x,i}$ s and the $\varepsilon_{x,y,ji}$ s. The output set $\{\beta_{on};\beta_{off}\}_{MC}$ is then used as an initial condition in the optimization procedure. Preliminary results showed that the set $\{\beta_{on};\beta_{off}\}_{MC}$ is favorable over the set $\{\beta_{on};\beta_{off}\}$ with respect to the sum of the differences, $\varepsilon_{x,i}$ s and $\varepsilon_{x,y,ji}$ s, in the cases tested.

Finally, we stress that the main aim of our algorithm, and the associated toolbox-methods, is to obtain a reliable unique mechanism only from the information contained in the data (namely, without relying on any preliminary information). As simple tests for any optimization subroutine presented here, the optimization is initialized near the analytical result. In all these tests, the analytical solution was obtained immediately.

## IV. SUMMARY AND CONCLUDING REMARKS

In this paper, we gave a toolbox for analyzing two-state trajectories generated by KSs. Our toolbox builds a canonical form of reduced dimensions (RD) based only on the information it extracts from the data. In the current context, a canonical form is a mechanism that is uniquely obtained from the data, and any *on-off* KS is equivalent to only one canonical form. Important features of RD forms include: RD forms are *on-off* networks with connections only between substates of different states. The connections have (usually) non-exponential WT-PDFs. A RD form has the simplest topology that can reproduce the data, but the WT-PDFs for the connections are usually multi-exponentials, and this reflects the complex topology of the underlying KS.

The toolbox executes a four-step algorithm:

(1) Estimation of the spectrum and amplitudes of the WT-PDFs of the single periods in the trajectory, $\phi_x(t)$ s: $\phi_x(t) = \sum_{i=1}^{L_x} c_{x,i} e^{-\lambda_{x,i} t}$ .

(2) Estimation of the matrices $\sigma_{x,y}$ s that appear in the exponential expansions of the $\phi_{x,y}(t_1,t_2)$ s: $\phi_{x,y}(t_1,t_2)=\sum_{i=1}^{L_x}\sum_{j=1}^{L_y}\sigma_{x,y,ij}e^{-\lambda_{x,i}t_1-\lambda_{y,j}t_2}$ .

(3) Determination of the numbers of substates in the two states in the RD form from the estimated ranks, $R_{x,y}$, of the $\phi_{x,y}(t_1,t_2)$ s. Mathematically, $R_{x,y}$ is the rank of $\sigma_{x,y}$ .

(4) Determination of the RD form starting from a low-resolution RD form constructed from the information collected in the first stages of the analysis. This step uses optimization subroutines for determining the pre-exponential coefficients, $\alpha_{x,jki}$, of the WT-PDFs for the connections in the RD form, $\varphi_{x,ji}(t)=\sum_{k=1}^{L_x}\alpha_{x,jki}e^{-\lambda_{x,k}t}$ .

Our toolbox is based on several new approaches for analyzing the *on-off* trajectory: firstly, it builds a canonical form directly from the data. Then, it builds the matrices $\sigma_{x,y}$, estimates the ranks of $\phi_{x,y}(t_1,t_2)$, and phrases the first mechanism-dependent optimization step as a root-search problem. To the best of our knowledge, none of these techniques were used before in the analysis of two-state trajectories, but, here, were proven crucial for the success of the analysis, i.e. constructing a reliable canonical form of reduced dimensions. The toolbox compiles, self-consistently, the information obtained from the various steps of the analysis for building the RD form. Our toolbox gives more accurate results than existing methods in the cases examined by us. The methods that constitute the backbone of our toolbox include:

(1) WT-PDFs with single arguments (i. e. $\phi_x(t)$, $\phi_{x+y}(t)$, $\phi_{\left(\sqrt{x}+\sqrt{y}\right)^2}(t)$, *x, y=on, off*) are found from the data by a subroutine that is based on the Padé approximation technique and a mechanism-free maximum likelihood procedure.

(2) The matrices $\sigma_{x,y}$ s, *x, y=on, off*, are estimated by a method that builds a matrix from the single argument WT-PDFs' amplitudes and rates (i.e. the expansion coefficients of $\phi_x(t)$, $\phi_{x+y}(t)$, $\phi_{\left(\sqrt{x}+\sqrt{y}\right)^2}(t)$, *x, y=on, off*), and uses this matrix in a numerical algorithm for estimating the matrices $\sigma_{x,y}$ s. A mechanism-free maximum likelihood procedure is applied as a final step.

(3) The ranks $R_{x,y}$ s of the $\phi_{x,y}(t_1,t_2)$ s are estimated by a method that analyzes the ratio of successive singular values in the decomposition of the *cumulative* WT-PDF of the *second order*. Crucial for the success of this technique is use of the *second order* cumulative WT-PDF to reduce noise, and the tuning of the bin-size of the 2D histogram by analyzing the randomized data.

(4) The coefficients in the exponential expansions of the $\varphi_{x,ij}(t)$ s are found by two-step optimization subroutines. The first optimization subroutine uses the coefficients extracted in the first steps of the analysis, but not the random times and is built as a root-search problem. The equations use the coefficients collected in the first stages of the analysis, and the physical requirements, $\varphi_{x,ij}(t) \geq 0$, for every *x*, *i* and *j*, phrased as demands on the couple first moments of the $\varphi_{x,ij}(t)$ s. The initial conditions in this optimization subroutine are found by an iterative algorithm with random updating. The output of this optimization can be used as an initial condition in a maximum likelihood subroutine with the random times. Expressions for the analytical derivatives of the target functions were derived (appendix C), and can be used in the optimization subroutines.

All the methods in our toolbox are automated in Matlab codes for a convenient use in the analysis of experimental data. Note that the first stages in our routine are flexible enough to be easily implemented in any type of routine, where the specificity of building the RD form from the data enters only in mature stages of the routine. We state that the toolbox is freely available for academic research upon electronic request.

Lastly, we briefly refer to the analysis of the particular example of protein's dynamics, e.g. (17). In these experiments, it is possible to translate the data into a two-state time-series, which means defining a folded state and an unfolded state. Each of these states can contain many semi-stable structures. The ensemble of structures constitutes the substates in the underlying KS. Our analysis of such a system gives the number of semi-stable structures in the folded and unfolded states, the number of structurally-different pathways that connect folded and unfolded structures (these are the numbers of substates in the RD form), and the dynamics of each pathway (these are the WT-PDFs for the connections in the RD form). It is important to note that this is the maximal amount of information that can be known on the dynamics of the protein from the folded-unfolded trajectory.

## ACKNOWLEDGMENTS

The authors acknowledge support from the NSF under grant CHE 0556268.

**Appendix A – Obtaining WT-PDFs from the data: Exponential fit subroutine**

This appendix presents our exponential fit subroutine. We assume that the experimental curve, $\phi_{\exp}(t)$, obeys an exponential expansion,

$$\phi_{\exp}(t) = \sum_{i=1}^{L} c_i e^{-\lambda_i t} , \tag{A1}$$

and aim at finding the coefficients, $c_i$s, the eigenvalues, $\lambda_i$s, and the optimal number of terms in the expansion, $L$. Note that $\phi_{\exp}(t)$ is defined for times fulfilling, $dt \le t \le T$. Here, $dt$ is the trajectory bin size and $T$ is the largest observed random time. The subroutine first smoothes the experimental curve according to (65-66),

$$\tilde{\phi}_{\exp}(t) = \frac{1}{T_d}\int_0^{T_d} \frac{\phi_{\exp}(t+q)}{\phi_{\exp}(q)} dq .$$

Here, $T_d$ determines the smoothing degree. When $T_d$ is small, the degree of smoothing is small. In the routine, we vary $T_d$ around $T_d \approx 0.05T$. (Note that for peaked input curve, $\phi_{\exp}(q)$ in the integrand's denominator should be replaced by $\phi_{\exp}(t_{\max} + q)$, where $t_{\max}$ is the time for which $\phi_{\exp}(q)$ is peaked.) It is straightforward to see that for the above smoothing procedure, $\tilde{\phi}_{\exp}(t)$ follows an exponential expansion with the same eigenvalues as of $\phi_{\exp}(t)$,

$$\tilde{\phi}_{\exp}(t) = \sum_{i=1}^{L} \tilde{c}_i e^{-\lambda_i t} ,$$

but with different coefficients $\tilde{c}_i$ s that are proportional to the original coefficients and obey,

$$\tilde{c}_i = \frac{c_i}{T_d}\int_0^{T_d} \frac{e^{-\lambda_i q}}{\phi_{\exp}(q)} dq .$$

The smoothing is performed to improve the accuracy of the numerical procedure, and after the smoothing step, we end up with the same problem of determining the coefficients, eigenvalues, and optimal order in an exponential expansion of an experimental curve. To continue, the Padé approximation technique is used (60-64). Firstly, we calculate, numerically, the averages of $t^n e^{-s_0 t}$ using the experimental curve [hereafter, we use the initial WT-PDF in Eq. (A1)],

$$mo_n(s_0) = \int_{dt}^{T} \phi_{\exp}(t) t^n e^{-s_0 t} dt \quad ; \qquad n = 1,...,2P . \tag{A2}$$

The set $\{mo(s_0)\}$ in Eq. (A2) contains the first $2P$ coefficients in a Taylor expansion of $\bar{\phi}_{\exp}(s)$ around $s_0$. In Eq. (A2), $P$ can be taken as large as needed, given that the numerical integral in Eq. (A2) converges for any $n$. The convergence of the integral for large values of $n$ can be achieved when $s_0$ is large. However, when $s_0$ becomes too large, $mo_n(s_0)$ for small values of $n$ is not accurately computed. Thus, $s_0$ should be chosen such that it gives the best fit, and this happens when the 'important' coefficients are of the same order of magnitude (60-61). It was suggested that the iterative formula (63),

$$s_{0,new} = s_{0,old} + 1 + mo_i(s_{0,old}) / mo_{i+1}(s_{0,old}),$$

where $i+1$ is the largest coefficient's order in the calculation, leads to this property. (In fact, under the mild assumptions that are automatically fulfilled by the acceptance of Eq. (A1), the iterative formula for $s_{0,new}$ leads to a ratio of unity for the largest (computed) successive moments.)

We use the moments $\{mo(s_0)\}$ in the Longman recursion relations (62), to get the Laplace transform expansion of $\phi_{\exp}(t)$. The Longman recursion relation gives the coefficients in a polynomial over polynomial expansion of a function, given its first $p$ Taylor coefficients, and this is exactly the problem at hand. In particular, the Longman recursion relations transform the information contained in the $\{mo(s_0)\}$ into a different set of coefficients $\{A, B\}$, such that,

$$\bar{\phi}_{\exp}(s) = \frac{A_0 + A_1(s - s_0) + A_2(s - s_0)^2 + ... + A_{p-1}(s - s_0)^{p-1}}{B_0 + B_1(s - s_0) + B_2(s - s_0)^2 + ... + B_p(s - s_0)^p}, \tag{A3}$$

where $B_0$ is set to one. (For details about the algorithm by Longman, see Ref. (62).) Equation (A3) is the exact Laplace transform of the function in Eq.(A1), but it also has the structure of a polynomial over a polynomial, thus, Eq. (A3) is the Padé representation of $\bar{\phi}_{\exp}(s)$ for $p=L$, around $s_0$. (If the original function in the time domain cannot be expressed analytically as a sum of exponentials, or if $p < L$ in Eq. (A3), than Eq. (A3) approximates the Laplace transform of the original function, but otherwise Eq.(A3) is exactly the Laplace transform of the original curve.) Thus, the Longman recursion relation gives the coefficients $\{A, B\}$ in Eq. (A3), from which the $c_i$s and $\lambda_i$s are computed by writing the polynomials in Eq. (A3) in root forms ($\xi = s - s_0$),

$$\bar{\phi}_{\exp}(s) = \frac{\prod_{i=1}^{p-1}(\xi - \tilde{A}_i)}{\prod_{i=1}^{p}(\xi - \tilde{B}_i)}, \tag{A4}$$

so,

$$\lambda_i = s_0 + \widetilde{B}_i \,, \tag{A5}$$

and,

$$c_i = \left( \frac{\prod_{j=1}^{p-1} (\xi - \widetilde{A}_j)}{\prod_{j \neq i}^{p} (\xi - \widetilde{B}_j)} \right)_{\xi = \widetilde{B}_i} \,. \tag{A6}$$

One of the advantages of using the Longman recursion relation is that for a given smoothing degree ($T_d$), and a given value of $s_0$, we get very cheaply, computationally-wise, a set of approximations to an increasing order in $p$ in Eq. (A3). For analytical curves, one can choose the best approximation among the obtained approximations by a stability analysis of the poles of residues of each approximation (60-61). We have found that the cheapest and most reliable way to choose the best approximation for a given experimental curve (and a given smoothing degree and a given value of $s_0$) is by using a non-linear mean least square criteria, which demands that the quantity,

$$\Delta x^2 = \sum_t \left( f_{fit}(t \mid \Theta) - \phi_{\exp}(t) \right)^2 ,$$

is minimal. Here, the function $f_{fit}(t \mid \Theta_p)$ is the fit function with the parameters $\Theta_p = \{A, B\}_p$ for order $p$ (namely, the parameters of the fit are calculated with the first $2p$ Taylor coefficients of the input curve).

We define the best 'local' fit as the fit that minimizes $\Delta x^2$ for a given smoothing degree and a given value of $s_0$. The comparison among the best 'local' fits, which are obtained while changing the smoothing degree and the value of $s_0$, is done by demanding a maximum for the likelihood score of the fit. Namely, for each 'local' best fit, we calculate the likelihood of the fit given the random times,

$$L_{loc} = \tfrac{1}{N_{tra}} \sum_{i=1}^{N_{tra}} \log( f_{fit}(t_i \mid \Theta_{loc})) ,$$

where $N_{tra}$ is the number of *on-off* cycles in the trajectory. (For large data set, we take only a fraction of the total data in the likelihood calculations). The best global fit is the one with the highest likelihood score, but we must compensate on the increase in the likelihood for fits with more components. This is done by using the Bayesian information criteria (BIC). BIC is formulated as, $BIC = -2L_{loc} + N_{\Theta} \frac{\log(N_{tra})}{N_{tra}}$, where $N_{\Theta}$ is the number of free parameters. The BIC score is minimal for the best fit.

The likelihood approach enables us also to change the criteria for finding 'local' best fits. For this we use, $\Delta x^2 = \sum_t \left( f_{fit}(t \mid \Theta)^{pow} - \phi_{\exp}(t)^{pow} \right)^2$, and in the routine we usually take, $pow = 1, 0.1, 0.01$ (a relative low *pow* value give a higher weight for small $f_{fit}(t \mid \Theta)$ values.).

The last step in our exponential fit subroutine directly maximizes the likelihood of the fit with the initial conditions being the best 'global' fit, and with constraints ensuring that the obtained fit is positive definite. For the direct maximization of the likelihood function we use the command '*fmincon*' in Matlab (so, in fact, we minimize minus log-likelihood). We note that the final step of direct maximization of the likelihood function doesn't change the number of components (the most prominent effect in this regard can be that some amplitudes become very small relative to others, or some rates become very close to each other), and, as usually happens with numerical minimizations, the final solution is sensitive to the initial condition. For our current purposes, this behavior is favorable, because the initial condition is already a very good solution for the problem. However, the optimization step can improve the results (up to 20% in some of the cases studied here), but cannot lead to a worse solution.

**Appendix B – Correlation analysis of the data**

This appendix is complementary to subsections III.2.1- III.2.2 in the main text, and gives our subroutines for finding and quantifying correlations in data. The first subsection in this appendix discusses the technical details for evaluating the results from the tests that check the existence of correlations in the data. The second subsection presents all the details of the subroutine that finds the binned- and generalized-binned successive WT-PDFs and the matrices $\sigma_{x,y}$s. The third subsection gives the details of the method that estimates the rank of an experimental 2D WT-PDF. The outputs of these three subroutines are estimations for the degree of correlations between events in the data and the matrices $\sigma_{x,y}$s.

**B.1 – Determining correlations between events** To use the methods of section **III.2.1** for detecting correlations in the data, we employ a standard statistical confidence test. We exemplify this test by evaluating the results from the moment analysis (method **a.3** of **III.2.1**). Say that we compute the average, $\left\langle t_{x,1} t_{y,2} \right\rangle = \frac{1}{N_{tra}} \sum_{i=1}^{N_{tra}} t_{x,i} t_{y,\delta_{x,y}+i}$ ,

from the data. To know whether the data is correlated or not, we compare the average of the same quantity obtained from the randomized data. In particular, we define the ratio,

$r_{M_{x,y,n}} = \left\langle t_{x,1_{x,1}}^{n} t_{y,2}^{n} \right\rangle / \left( \left\langle t_x \right\rangle \left\langle t_y \right\rangle \right)^n$ .

The first order difference uses $r_{M_{x,y,1}}$ ,

$D = \left| r_{M_{x,y,1}} - 1 \right| \geq 0$ .

The significance-test determines the upper bound for which $D$ is still regarded as zero, meaning that the data is uncorrelated. To find this upper bound, we associate errors with the moment values. The error in $\left\langle t_{x,1} t_{y,2} \right\rangle$ is obtained by dividing the trajectory into segments of $N_{eve}$ events (so there are $N_{seg} = N_{tra} / N_{eve}$ segments), and calculating the quantity,

$$\varepsilon = \sqrt{\left\langle \left\langle t_{x,1} t_{y,2} \right\rangle^2 \right\rangle_{O.S.} - \left\langle \left\langle t_{x,1} t_{y,2} \right\rangle \right\rangle_{O.S.}^2} = \left\langle t_{x,1} t_{y,2} \right\rangle \sqrt{\left\langle \left\langle t_{x,1} t_{y,2} \right\rangle^2 \right\rangle_{O.S.} \Big/ \left\langle t_{x,1} t_{y,2} \right\rangle^2 - 1} .$$

Here, $\left\langle \bullet \right\rangle_{O.S.}$ averages the argument $\bullet$ over the segments, and the equality, $\left\langle \left\langle t_{x,1} t_{y,2} \right\rangle \right\rangle = \left\langle t_{x,1} t_{y,2} \right\rangle$ was applied to get the final expression. $\left\langle \left\langle t_{x,1} t_{y,2} \right\rangle^2 \right\rangle_{O.S.}$ is given by,

$$\left\langle \left\langle t_{x,1} t_{y,2} \right\rangle^2 \right\rangle_{O.S.} = \frac{1}{N_{seg}} \sum_{i=1}^{N_{seg}} \left\langle t_{x,1} t_{y,2} \right\rangle_i^2 ,$$

and, in particular,

$$\left\langle \left\langle t_{x,1} t_{y,2} \right\rangle^2 \right\rangle_{O.S.} = \frac{1}{N_{seg}} \sum_{i=1}^{N_{seg}} \left( \frac{1}{N_{eve}} \sum_{j=1}^{N_{eve}} t_{x,j}^{i} t_{y,j+\delta_{x,y}}^{i} \right)^2$$

$$= \frac{1}{N_{seg}} \left( \frac{1}{N_{eve}} \right)^2 \sum_{i=1}^{N_{seg}} \sum_{j=1}^{N_{eve}} t_{x,j}^{i} t_{y,j+\delta_{x,y}}^{i} \sum_{k=1}^{N_{eve}} t_{x,k}^{i} t_{y,k+\delta_{x,y}}^{i}$$

$$= \left( \frac{1}{N_{eve}} \right)^2 \sum_{j,k=1}^{N_{eve}} \left\langle t_{x,j} t_{y,j+\delta_{x,y}} t_{x,k} t_{y,k+\delta_{x,y}} \right\rangle$$

$$= \frac{1}{N_{eve}} \left( \left\langle t_{x,1}^2 t_{y,2}^2 \right\rangle + \left\langle t_{x,1} t_{y,2} t_{x,2} t_{y,3} \right\rangle + \ldots \right).$$

The sum in the last line has $N_{eve}$ terms. Excluding the first term, all terms approximately equal to $\left\langle t_{x,1} t_{y,2} \right\rangle^2$ , and thus, for large $N_{eve}$, we get,

$$\left\langle \left\langle t_{x,1}^2 t_{y,2}^2 \right\rangle^2 \right\rangle_{O.S.} \approx \frac{1}{N_{eve}} \left\langle t_{x,1}^2 t_{y,2}^2 \right\rangle + \left\langle t_{x,1} t_{y,2} \right\rangle^2 .$$

In terms of $\varepsilon$,

$$\varepsilon \equiv \langle t_{x,1} t_{y,2} \rangle \delta \approx \langle t_{x,1} t_{y,2} \rangle \sqrt{\langle t_{x,1}^2 t_{y,2}^2 \rangle \Big/ \langle t_{x,1} t_{y,2} \rangle^2} \sqrt{\tfrac{1}{N_{eve}}} .$$

Similar analysis is performed on the randomized data, and leads to,

$$\delta_{N.C.} = \sqrt{\left\langle \langle t_x \rangle^2 \langle t_y \rangle^2 \right\rangle_{O.S.} \Big/ \langle t_x \rangle^2 \langle t_y \rangle^2 - 1} \approx \sqrt{\tfrac{1}{N_{eve}}} \sqrt{\langle t_x^2 \rangle \langle t_y^2 \rangle} \Big/ \langle t_x \rangle \langle t_y \rangle .$$

Note that in practice, the averages are calculated with the entire data, and $N_{eve} = N_{tra}$. Finally, we look on the quantity,

$$P_{CR} = D - \alpha\sigma \qquad ; \qquad \sigma = r_{M_{x,y,1}} \delta + \delta_{N.C.} .$$

When $\alpha = 1$ and $P_{CR} \geq 0$, there are correlations in the data with 85% of confidence. Increasing the value of $\alpha$ while keeping $P_{CR} \geq 0$, increases the confidence level of the result. For example, for $\alpha = 1.39$ and $P_{CR} \geq 0$, there are correlations in the data with 95% of confidence, and for $\alpha = 3$ with $P_{CR} \geq 0$ there are correlations in the data with 99.998% of confidence. The confidence values assume Gaussian noise (around the means). For the current problem, this is an adequate assumption for large $N_{eve}$.

The same analysis done above for $\langle t_{x,1} t_{y,2} \rangle$ is performed also to higher order moments of the two-dimensional WT-PDF of successive events. For example, for,

$$\langle t_{x,1}^2 t_{y,2}^2 \rangle = \tfrac{1}{N_{tra}} \sum_{i=1}^{N_{tra}} t_{x,i}^2 t_{y,i+\delta_{x,y}}^2 ,$$

the error factor is given by,

$$\delta = \sqrt{\left\langle \langle t_{x,1}^2 t_{y,2}^2 \rangle^2 \right\rangle_{O.S.} \Big/ \langle t_{x,1}^2 t_{y,2}^2 \rangle^2 - 1} \approx \sqrt{\tfrac{1}{N_{eve}}} \sqrt{\langle t_{x,1}^4 t_{y,2}^4 \rangle \Big/ \langle t_{x,1}^2 t_{y,2}^2 \rangle^2} .$$

**B.2 – The matrices $\sigma_{x,y}$ s** This subsection estimates the matrices $\sigma_{x,y}$ s. For example, the matrix $\sigma_{on,off}$, with dimensions $L_{on}$,$L_{off}$, gives $\phi_{on,off}(t_1,t_2)$,

$$\phi_{on,off}(t_1,t_2) = \sum_{i,j=1}^{L_{on} . L_{off}} \sigma_{on,off,ij} e^{-t_2 \lambda_{y,j} - t_1 \lambda_{x,i}} ,$$

and consequently determines the binned successive WT-PDF,

$$\phi_{on+off}(t) = \sum_{i,j} \sigma_{on,off,ij} (e^{-\lambda_{on,i} t} - e^{-\lambda_{off,j} t}) / (\lambda_{off,j} - \lambda_{on,i}) ,$$

and the generalized-binned successive WT-PDFs,

$$\phi_{(\sqrt{on}+\sqrt{off})^2}(t) \Big/ t/3 \approx \sum_{i=1}^{L_{on}} \sum_{j=1}^{L_{off}} \sigma_{on,off,ij} e^{-0.3t(\lambda_{on,i} + \lambda_{off,j})} .$$

(The generalized-binned successive WT-PDFs given above is in the short time approximation, $t(\lambda_{on,i} + \lambda_{off,j}) < 1$ for every *i* and *j*.)

In our routine, we build $\phi_{x+y}(t)$ from the trajectory, and use the exponential fit command to find its exponential expansion. The *amplitudes* of $\phi_{x+y}(t)$ are found by matching the rates obtained from the fit (denoted by $\lambda_{x+y}$) and those found for the $\phi_x(t)$s (namely, the set $\{\lambda_x, \lambda_y\}$). Then, a mechanism-free maximum likelihood procedure is performed. Here, the rates $\{\lambda_x, \lambda_y\}$ are input constants, the amplitudes $\sigma_{x+y,i}$s are the variables, initialized by the values found by the fit and the rate-matching procedure.

Note that, usually, not all the $L_x$+$L_y$ components are recovered from the fit of $\phi_{x+y}(t)$. To estimate the missing amplitudes, the rate-matching procedure is applied for all the entries in the spectrum $\{\lambda_x, \lambda_y\}$ using the fit rates $\lambda_{x+y,i}$s. This artificially creates a situation in which different rates have the same amplitudes in the expression for $\phi_{x+y}(t)$, and it is corrected by dividing the amplitudes of the same $\lambda_{x+y,i}$ according to the order of appearance.

To get the matrices $\sigma_{x,y}$s, we start with fitting $\phi_{\left(\sqrt{x}+\sqrt{y}\right)^2}(t) \big/ t/3$ to a sum of exponentials, with as many as $L_xL_y$ terms, using the exponential fit subroutine. Then, matching a fit rate to $0.3(\lambda_{x,i} + \lambda_{y,j})$ gives initial value for entries in the matrix $\sigma_{x,y}$. These are the initial values for the variables in a mechanism-free maximum likelihood procedure, where the target function is built from $\phi_{x,y}(t_1,t_2)$.

Similar to the note given above for the construction of the $\sigma_{x+y,i}$s, the number of terms found from the fit of $\phi_{\left(\sqrt{x}+\sqrt{y}\right)^2}(t) \big/ t/3$ is usually smaller than the number of the elements in $\sigma_{x,y}$ ($\equiv L_xL_y$). To estimate the missing entries, a matrix equation relates the $\sigma_{x,y,ji}$s to all the coefficients that were already found in the routine. (The discussion below is made for particular *x* and *y* values, *x*=*on* and *y*=*off*, but the same operations are done for any other combination of *x* and *y*). In particular, we define a vector $\vec{V}_{\sigma_{on.off}}$ such that,

$$\left(\vec{V}_{\sigma_{on.off}}\right)_{i+L_{on}(j-1)} = \sigma_{on,off,ij}$$

and a matrix *Ac*, with dimensions $\lfloor 2(L_{on}+L_{off}), L_{on}L_{off} \rfloor$, and entries,

$$(Ac)_{ij} = \begin{cases} 1/\lambda_{n,k} & 1 \le i \le L_{off}; k = j + (i-1)L_{on} \\ 1/\left(\lambda_{f,k} - \lambda_{n,i-L_{off}}\right) & 1 + L_{off} \le i \le 2L_{off}; k = j + (i-1-L_{off})L_{on} \\ 1/\lambda_{n,k} & 1 + 2L_{off} \le i \le 2L_{off} + L_{on}; k = (j-1)L_{on} + i - 2L_{off} \\ 1/\left(\lambda_{n,k} - \lambda_{f,i-2L_{off}+L_{on}}\right) & 1 + 2L_{off} + L_{on} \le i \le 2(L_{off}+L_{on}); k = (j-1)L_{on} + i - 2L_{off} \\ 0 & otherwise \end{cases}$$

The product matrix-vector gives the fit amplitudes,

$$Ac\vec{V}_{\sigma_{on.off}} = \vec{V}_c .$$

Here, $\vec{V}_c$ is given by,

$$\left(\vec{V}_c\right)_i = \begin{cases} c_{off,i} & 1 \le i \le L_{off} \\ \sigma_{on+off,i-L_{off}} & 1 + L_{off} \le i \le 2L_{off} \\ c_{off,i-2L_{off}} & 1 + 2L_{off} \le i \le 2L_{off} + L_{on} \\ \sigma_{on+off,i-2L_{off}-L_{on}} & 1 + 2L_{off} + L_{on} \le i \le 2(L_{off}+L_{on}) \end{cases}$$

If only some of the elements of $\vec{V}_{\sigma_{on.off}}$ can be estimated from the fit of $\phi_{\left(\sqrt{on}+\sqrt{off}\right)^2}(t)\big/t/3$, only the rest can be estimated by the above matrix operation. The maximal number of unknowns that can be obtained from this approach is exactly the rank of matrix *Ac*, where the rank cannot exceed the minimum between $2(L_{on}+L_{off})$ and $L_{on}L_{off}$. We note that the number of rows in the matrix *Ac* can be extended by building from the data the WT-PDF $\phi_{|x-y|}(t)$, which may, in some cases, lead to an increase in the rank of matrix *Ac*, and to better results. In any case, when $L_x \le 2$, the number of rows in always smaller than the number of unknowns without extending the number of rows in matrix *Ac*.

Usually the matrix *Ac* is not a square matrix, so we use its singular value decomposition to obtain the unknowns: let,

$$Ac = USV',$$

be the singular value decomposition of matrix *Ac*, then

$$\vec{V}_{\sigma_{on.off}} = VS^{-1}U'\vec{V}_c .$$

Here, we use the property, $VV' = 1; UU' = 1$, and take $S$ a square matrix. To use the information obtained from the fit of $\phi_{\left(\sqrt{on}+\sqrt{off}\right)^2}(t)\big/t/3$, we write $\vec{V}_{\sigma_{on.off}} = \vec{V}^K{}_{\sigma_{on.off}} + \vec{V}^U{}_{\sigma_{on.off}}$ (the superscript $K$ stands for known elements found from the fit of $\phi_{\left(\sqrt{on}+\sqrt{off}\right)^2}(t)\Big/t/3$ and the matching procedure, and the superscript *U* stands for unknown elements), and modified vector $\vec{V}_c$ to, $\vec{V}_c^M = \vec{V}_c - Ac\vec{V}^K{}_{\sigma_{on,off}}$, such that, $\vec{V}^U{}_{\sigma_{on.off}} = VS^{-1}U'\vec{V}_c^M$.

Theoretically, when the rank of matrix *Ac* equals to the number of unknown, it is sufficient to apply the matrix operation, without a modification of the vector $\vec{V}_c$, to get the unknowns. However, as long as the rank of matrix *Ac* equal the number of elements in the vector $\vec{V}^U{}_{\sigma_{on.off}}$, it is possible to get all the unknowns from the above method (where vector $\vec{V}_c$ is modified). As, the elements in the matrix *Ac* contain statistical errors, a mechanism-free maximum likelihood procedure is done to get the final estimation for $\vec{V}_{\sigma_{on.off}}$, with initial condition being the elements of the vector $\vec{V}_{\sigma_{on.off}}$ found from the matrix operation. We can produce many different sets of initial conditions, by choosing different elements in the sets $\vec{V}^K{}_{\sigma_{on.off}}$ and $\vec{V}^U{}_{\sigma_{on.off}}$. The best fit maximizes the likelihood function.

**B.3 Rank estimation** This subsection gives the subroutine for estimating the rank of a 2D histogram. Technically, we work with a fifty by fifty matrix, with an initial bin size *dt*. The final bin size is determined by performing the subroutine described below on the *randomized data*, and demanding that the resulting rank is one. This reference-point-method for determining the actual bin size in the calculation of the rank of $\phi_{x,y}(t_1,t_2)$ is found important for achieving accurate results.

The calculation starts with the construction of the *second* order cumulative histogram, and obtains its spectrum of singular values. The ratio of successive singular values is examined; the rank equals to the order of the ratio that doesn't exceed a predetermined threshold value for the first time, minus one.

The accuracy of the result can be improved by looking on the value of largest noise singular value, if we can estimate the noise in the histogram. It was shown in (69), that for a sum of a rank *r* matrix of dimensions [*m,n*] with a full rank (Gaussian) noise matrix of variance $\sigma$, the singular value of order *r+1* lies within the boundaries,

$$\sigma\sqrt{c} \le \lambda_{r+1} \le \sigma\sqrt{mn}$$

where $c$ is determined by the user-defined significance-level obtained from a $\chi^2$ distribution with $m$ degrees of freedom. These boundaries assume that the $r^{th}$ singular value of the unperturbed matrix is, at least, twice as large as the largest singular value of the noise matrix. Namely, the noise must be small for the ratio method to work. We have found that the second order cumulative 2D histogram works best for our purposes in reducing the noise. We note that an estimation of the bounds of the noise level in the 2D cumulative histogram of order $n$ can be obtained by the equation,

$$\sqrt{\left\langle t_{x,1}^2 t_{y,2}^2 \right\rangle_{C_n} \Big/ N_{C_n}} \le \sigma \le \sqrt{\left\langle t_{x,1}^2 t_{y,2}^2 \right\rangle_{C_n} \Big/ N_{bin}}$$

where $\left\langle t_{x,1}^2 t_{y,2}^2 \right\rangle_{C_n}$ is calculated with all events ($\equiv N_{C_n}$) in the 2D cumulative histogram of order $n$, and $N_{bin}$ is the number of events in the first bin of the cumulative histogram.

Finally, we estimate the threshold value for the ratio test by taking the ratio of the upper to lower bounds of $\lambda_{r+1}$,

$$threshold = \sqrt{mn} / \sqrt{c}\,.$$

For a fifty-by-fifty matrix the threshold value is 5.73 for significance level of 0.99.

**Appendix C – Finding the WT-PDFs for the connections in the RD form**

This appendix gives the mathematical details of the subroutine that builds the WT-PDFs for the connections in the RD form, and it is complementary to subsection III.3 in the main text.

We start by expressing $\phi_x(t)$ using the WT-PDFs for the connections in the RD form,

$$\phi_x(t) = \sum_{j=1}^{L_y} \sum_{i=1}^{L_x} W_{x,i} \varphi_{x,ji}(t)\,. \qquad \text{(C1)}$$

$\varphi_{x,ji}(t)$ in Eq. (C1) is the WT-PDF for connecting substate $i$ in state $x$ to substate $j$ in state $y$ in the RD form, and is given by,

$$\varphi_{x,ji}(t) = \sum_{k=1}^{L_x} \alpha_{x,jki} e^{-\lambda_{x,k} t}\,. \qquad \text{(C2)}$$

Equation (C2) introduces a set of parameters $\{\alpha_{on};\alpha_{off}\}$. The aim is to determine these parameters from the data for a complete specification of the RD form. (Recall that at this stage, the numbers of substates in both states in the RD form are known from the analysis of the previous stages, as well as the eigenvalues, and the order of the expansion in Eq. (C1).) To estimate the set $\{\alpha_{on};\alpha_{off}\}$, the $W_{x,i}$ s in Eq. (C1) should be related to the $\{\alpha_{on};\alpha_{off}\}$. The $W_{x,i}$ s are the normalized steady state fluxes from state $y$ to substate $i$ in state $x$, and are completely defined by the zeroth and first order moments of the $\varphi_{x,ji}(t)s$. (Note that the zeroth order moment of $\varphi_{x,ji}(t)$ is not unity if substate $i_x$ has more than one ongoing connection. Particularly, $\omega_{x,ji} = \int_0^\infty \varphi_{x,ji}(t)dt$ is the probability for the transition, $i_x \to j_y$.) The technical details for relating the normalized steady state fluxes to the $\{\alpha_{on};\alpha_{off}\}$ are given in the next subsection.

The calculation of $\{\alpha_{on};\alpha_{off}\}$ is carried out in two steps. First, a corresponding set of coefficients, denoted by $\{\beta_{on};\beta_{off}\}$, is calculated. The set $\{\beta_{on};\beta_{off}\}$ is found in an iterative algorithm that has a random updating role. We run $4\left((L_{on}+L_{on})L_{RD,on}L_{RD,on}\right)^2$ iterations, and save the best result (the best result minimizes the distance between the RD forms $\sigma_{x,y}$ s and its actual experimental values). The best set is the $\{\beta_{on};\beta_{off}\}$. The second step in the construction of the $\{\alpha_{on};\alpha_{off}\}$ uses the $\{\beta_{on};\beta_{off}\}$ as an initial condition in an optimization subroutine with constraints and analytical derivatives. Now, the set $\{\beta_{on};\beta_{off}\}$ is not unique (in fact, there are infinitely many sets $\{\beta_{on};\beta_{off}\}$, due to the non-linearity of the iterative equations), and the parameter space is non-continuous (this is a consequence of the physical demands that the WT-PDFs must fulfill). Therefore, the optimization in the second stage of this subroutine usually (actually, almost always) finds local minimum. The solution to this difficulty is to produce many initial sets. The result $\{\alpha_{on};\alpha_{off}\}$ has the best final score in the likelihood calculations. We note that the averaging over initial sets $\{\beta_{on};\beta_{off}\}$ is the most time consuming step in our analysis, and can take half a day for a 30 parameter system on a standard PC with ~2.66 GHz processor and 1 GB RAM.

Technical details for performing the above subroutine are discussed in the subsequent subsections.

**The weights** $W_{x,i}$ To get the weights $W_{x,i}$ s in Eq.(C1), we start with the definition,

$$W_{x,i} = \sum_j J_{y\to x,ij} / \sum_{ij} J_{y\to x,ij}\,, \tag{C3}$$

where the flux $J_{y\to x,ij}$ obeys,

$$J_{y\to x,ij} = \Gamma_{y,ij} P_{y,j}(ss). \tag{C4}$$

In Eq. (C4), $\Gamma_{y,ij} = \frac{\omega_{y,ij}}{<t_{y,j}>}$, where $<t_{y,j}> = \sum_i \int_0^\infty t\varphi_{y,ij}(t)dt$. We also force the normalization, $1 = \sum_i \int_0^\infty \varphi_{y,ij}(t)dt$. (When mapping a KS into a RD form this normalization immediately follows, but needed to be enforced in the analysis of data from experiments.) Note that $\Gamma_{y,ij}$ is the *ij* element of matrix $\mathbf{\Gamma}_y$. Now, to get the $W_{x,i}$ s we need the $P_{y,j}(ss)$ s in Eq. (C4), which is the probability to occupy substate *j* in state *y* in the RD form at steady state. This probability is found from the steady state equation,

$$\mathbf{\Gamma}\vec{P}(ss) = 0. \tag{C5}$$

In Eq. (C5), $\vec{P} = \begin{pmatrix} \vec{P}_{on}(ss) \\ \vec{P}_{off}(ss) \end{pmatrix}$, and $P_{z,j}(ss) = \left(\vec{P}_z(ss)\right)_j$. Matrix $\mathbf{\Gamma}$ is defined by,

$$\mathbf{\Gamma} = \begin{pmatrix} -diag\left(\vec{\mathbf{1}}_{off}\mathbf{\Gamma}_{on}\right) & \mathbf{\Gamma}_{off} \\ \mathbf{\Gamma}_{on} & -diag\left(\vec{\mathbf{1}}_{on}\mathbf{\Gamma}_{off}\right) \end{pmatrix}, \tag{C6}$$

where the operation *diag* in Eq. (C6) takes a vector and produces from it a square diagonal matrix whose $i^{th}$ diagonal element is the $i^{th}$ element of the original vector. $\vec{P}(ss)$ is the normalized eigenvector of matrix $\mathbf{\Gamma}$ that corresponds to the zero eigenvalue, and is easily found numerically.

**The iterative scheme for producing the coefficients** $\{\beta_{on};\beta_{off}\}$ We start by introducing the analytical relationships between the coefficients of the $\phi_x(t)$ s and the $\{\alpha_{on};\alpha_{off}\}$,

$$c_{x,H} = \sum_{j=1}^{L_{RD,y}} \sum_{i=1}^{L_{RD,x}} W_{x,i}\alpha_{x,jHi} \quad ; \quad H = 1,...,L_x. \tag{C7}$$

Equations (C7), $L_x$ in number, are obtained by comparing Eq. (2) and Eq. (C1) after substituting Eq. (C2) into it. Similarly, the amplitudes $\sigma_{x,y,ji}$ can be analytically related to the $\{\alpha_{on};\alpha_{off}\}$. For $x \neq y$ we have,

$$\sigma_{x,y,GH} = \sum_{j=1}^{L_{RD,y}} \sum_{i=1}^{L_{RD,x}} \sum_{k=1}^{L_{RD,y}} W_{x,i}\alpha_{x,jGi}\alpha_{y,kHj} \quad ; G = 1,...,L_x,\ H = 1,...,L_y, \tag{C8}$$

and for *x*=*y*,

$$\sigma_{x,x,GH} = \sum_{i=1}^{L_{RD,x}} \sum_{j=1}^{L_{RD,y}} \sum_{j'=1}^{L_{RD,x}} \sum_{k=1}^{L_{RD,y}} W_{x,i}\alpha_{x,jGi}\overline{\varphi}_{y,j'j}(0)\alpha_{x,kHj'} \quad ;$$

$$G = 1,...,L_x,\ H = 1,...,L_x. \tag{C9}$$

We use Eqs. (C7)-(C9) to update the $\{\beta_{on};\beta_{off}\}$,

$$\beta_{x,j'H'i'} = \beta_{x,j'H'i'} + \varepsilon_{x,H}, \tag{C10}$$

with $\varepsilon_{x,H} = c_{x,H} - \sum_{j=1}^{L_{RD,y}} \sum_{i=1}^{L_{RD,x}} W_{x,i}\beta_{x,jHi}$ being the difference between the actual value of $c_{x,H}$ and its approximation using the temporarily RD form, and,

$$\beta_{x,j'H'i'} = \beta_{x,j'H'i'} + \varepsilon_{x,y,HG} \frac{sign(VD_{y,Gj'})}{VP_{y,Gj'}}, \tag{C11}$$

where, $\varepsilon_{x,y,GH} = \sigma_{x,y,GH} - \sum_{j=1}^{L_{RD,y}} \sum_{i=1}^{L_{RD,x}} \sum_{k=1}^{L_{RD,x}} W_{x,i}\beta_{x,jGi}\beta_{y,kHj}$ $(x \neq y)$ is the difference between the actual value of $\sigma_{x,y,GH}$ and its approximation using the temporarily RD form. The sign of quantity $VD_{y,Gj} = \sum_{k=1}^{L_{RD,x}} \beta_{y,kGj}$ ensures the stability of the updating-scheme, as well as the quantity $VP_{y,Gj} = \sum_{k=1}^{L_{RD,x}} |\beta_{y,kGj}|$ (namely, these factors are introduced for decreasing the error in the RD form value of $\sigma_{x,y,GH}$ after the update).

The actual indices that determine the $\beta_{x,j'H'i'}$ to be updated are chosen randomly, but *on* and *off* coefficients are updated sequentially. After each iteration, the weights are updated. To initialize the iterative algorithm, a symmetric configuration is chosen: $W_{x,i} = 1/N_x$, and, $\beta_{x,jHi} = c_{x,H}/N_y$. The number of iterations is proportional to the square of the number of parameters. (There is no convergence role that stops the updating, but the best set along the iterations is saved.)

There are also several conditions that the $\{\beta_{on};\beta_{off}\}$ must fulfill. These are derived by demanding that the $\{\varphi_{on}(t);\varphi_{off}(t)\}$ are positive for every value of *t* in the time range of the experiment, namely,

$$\varphi_{x,ij}(t) \geq 0 \quad ; \quad dt \leq t \leq T, \tag{C12}$$

for every *i*, *j*, and *x*. (Every $\varphi_{x,ij}(t)$ decays to zero for sufficiently long time, by construction). When $\varphi_{x,ij}(t)$ becomes negative for a particular value of *t* that is relevant to the experiment, the most relevant negative coefficient (determined by its conjugated rate) is halved. If all the amplitudes are negative, they are all made positive. This sign inversion was found to have an important role in the success of the subroutine, because it also connects, otherwise unconnected, regions in the parameter space. When Eq.(C12) is satisfied for every *i*, *j*, and *x*, the normalization condition, $\sum_j \int_0^\infty \varphi_{x,ji}(t)dt = 1$, with $0 \leq \int_0^\infty \varphi_{x,ji}(t)dt \leq 1$ (for ensuring that $\int_0^\infty \varphi_{x,ji}(t)dt$ has the meaning of probability), can be easily obtained by an appropriate division. Note that the non-negativity condition with the appropriate normalization ensure that the probabilities of Eq.(C5) are all non-negative and do not exceed unity, namely, $\vec{0} \le \vec{P}(ss) \le \vec{1}$. In fact, to fulfill $\vec{0} \le \vec{P}(ss) \le \vec{1}$, we need to demand only that the zeroth and first moments of $\varphi_{x,ij}(t)$ are positive, for every *i, j, x*.

**The optimization subroutine for obtaining the** $\{\alpha_{on};\alpha_{off}\}$ Each set $\{\beta_{on};\beta_{off}\}$ is used as an initial condition in an optimization subroutine. The optimization subroutine finds a common root for a set of equations. The optimization uses the command '*fsolve*' in Matlab. The equations are the $\varepsilon_{x,H}$ s and $\varepsilon_{x,y,HG}$ s defined in the previous subsection, but there are also equations that guarantee the normalization of the $\varphi_{x,ij}(t)$ s. This is done by demanding that the *0*$^{th}$ and *n*$^{th}$ moments of each $\varphi_{x,ij}(t)$ are positive, with $\sum_i \int_0^\infty \varphi_{x,ij}(t)dt = 1$ for every *j,* and *x*. Note that when the *0*$^{th}$ and *n*$^{th}$ moments of $\varphi_{x,ij}(t)$ are positive, all the moments in between them are positive also *only* for a two-component $\varphi_{x,ij}(t)$ (higher moments than the *n*$^{th}$ one can always be negative), but we have found that this is a good-working condition also when $\varphi_{x,ij}(t)$ has more than two components. In this study, an extensive conditioning on the moments of $\varphi_{x,ij}(t)$ led to worse results (relative to a smaller number of constraints), because it restricts the available parameter space that the optimization subroutine can explore. The same is true for large values of *n*. In particular, the maximal number of moments of $\varphi_{x,ij}(t)$ that were constrained and led to good results is three: $1 \ge < t^0 > \ge 0$ , $< t^1 > \ge 0$, $< t^n > \ge 0$ with $4 \ge n > 1$, for every *i, j,* and *x*. *n*=*3* led to the best results and *n*=*5* completely damaged the search, and almost always (>99%) the optimization output was the initial point.

The above optimization subroutine does not use the actual data, but only the information that was extracted from it, which is stored in the coefficients $c_{x,H}$ s and the $\sigma_{x,y,HG}$ s. However, an additional step can use the data in a mechanism-dependent maximum likelihood subroutine, with the found $\{\alpha_{on};\alpha_{off}\}$ being the initial condition. The likelihood function reads,

$$l(\mathbf{\Theta}\,|\,data) = \tfrac{1}{4}\sum_{x,y} l_{x,y}(\mathbf{\Theta}\,|\,data)\,, \tag{C13}$$

where,

$$l_{x,y}(\mathbf{\Theta}\,|\,data) = \log[L_{x,y}(\mathbf{\Theta}\,|\,data)] = \tfrac{1}{N_{tra}-\delta_{x,y}} \sum_{j=1}^{N_{tra}-\delta_{x,y}} \log[\phi_{x,y}(t_{x,j}, t_{y,j+\delta_{x,y}})]\,.$$

In Eq. (C13), $\mathbf{\Theta}$ represents the set $\{\alpha_{on};\alpha_{off}\}$. The minimization procedure uses the command '*fmincom*' in Matlab. (The command can choose an algorithm from a variety of optimization procedures, but usually the algorithm used for the current problem is the quasi-Newton line search algorithm.) The constraints in the minimization subroutine demand $\varphi_{x,ij}(t) \geq 0$, for every *i*, *j*, and *x*, and relevant *t*. In both optimization subroutines, the analytical derivatives of the target function can be used, and the way to obtain them is given in the next subsection. We note that in all optimization subroutines, we first use numerical derivatives because this enables controlling the maximal change in the variable values. We have found that, in some cases, better results are obtained when allowing relatively large values for the maximal change in the variables. Note that the second optimization may improve the results when the $\sigma_{x,y,ij}$ s are relatively noisy (as can be deduced by an error analysis using the $\phi_{(\sqrt{x}+\sqrt{y})^2}(t)$ s).

**The analytical derivatives for the optimization subroutine** In our approach, the variables in the optimization subroutine are the $\{\beta_{on};\beta_{off}\}$. To get the analytical derivatives of the target function we first note that the $\varphi_{x,ji}(t)s$ are linear in the $\{\beta_{on};\beta_{off}\}$, and therefore the derivatives of the $\varphi_{x,ji}(t)s$ are easily performed. Taking the analytical derivatives of the $W_{x,i}$ s with respect to the $\{\beta_{on};\beta_{off}\}$ is harder. From Eqs. (C3)-(C4), we see that the difficulty in taking the derivatives of the $W_{x,i}$ s is to take the derivatives of the steady state probabilities of matrix $\mathbf{\Gamma}$, the $P_{z,j}(ss)$ s. The solution for this problem is to formulate it in a form of an exponential of a matrix, and to take the derivatives of an exponential of a matrix with respect to its entries. We proceed by recalling that the steady state probabilities of matrix $\mathbf{\Gamma}$ can be obtained from the mean residence times of a related matrix $\tilde{\mathbf{\Gamma}}$,

$$\tilde{\mathbf{\Gamma}} = \mathbf{\Gamma} - \varepsilon \cdot diag(1,0,0,0...0)\,.$$

Here, an irreversible transition from substate $1_{on}$ is added, and is of magnitude $\varepsilon$. $\varepsilon$ doesn't affect the final result, but for practical reasons it is taken to have the value of the overall decaying rate of substate $1_{on}$ in matrix $\mathbf{\Gamma}$. Matrix $\tilde{\mathbf{\Gamma}}$ doesn't have a steady-state solution, i.e. at infinite times the process is sure to occupy the added trap. However, the mean residence times of this system when starting at state $1_{on}$, which are the elements of the column vector, $\vec{\tau}(1)$, are proportional to the steady-state probabilities of matrix $\mathbf{\Gamma}$:

$$\vec{P}(ss) = \vec{\tau}(1) / \left(\vec{U}\vec{\tau}(1)\right).$$

Here, $\vec{U}$ is the summation row vector of appropriate dimensions, and vector $\vec{\tau}(1)$ is defined by,

$$\vec{\tau}(1) = \int_0^\infty e^{\tilde{\mathbf{\Gamma}}t} \vec{\tilde{P}}_0 dt = -\tilde{\mathbf{\Gamma}}^{-1} \vec{\tilde{P}}_0,$$

where $\vec{\tilde{P}}_0$ is the vector of the initial occupancies, $\left(\vec{\tilde{P}}_0\right)_j = \delta_{1j}$. The derivative of $\vec{\tau}(1)$ with respect to any element of matrix $\mathbf{\Gamma}$, denoted by $\gamma$, $\partial\vec{\tau}(1)/\partial\gamma$, is given by,

$$\partial\vec{\tau}(1)/\partial\gamma = \tfrac{\partial}{\partial\gamma} \int_0^\infty e^{\tilde{\mathbf{\Gamma}}t} \vec{\tilde{P}}_0 dt = \int_0^\infty \left(\tfrac{\partial}{\partial\gamma} e^{\tilde{\mathbf{\Gamma}}t}\right) \vec{\tilde{P}}_0 dt. \tag{C14}$$

Equation (C14) expresses $\partial\vec{\tau}(1)/\partial\gamma$ as a derivative of an exponential of a matrix, where all the eigenvalues of this matrix are negative, thus,

$$\partial\vec{\tau}(1)/\partial\gamma = \tilde{\mathbf{\Gamma}}^{-1} \left(\tfrac{\partial}{\partial\gamma} \tilde{\mathbf{\Gamma}}\right) \tilde{\mathbf{\Gamma}}^{-1} \vec{\tilde{P}}_0. \tag{C15}$$

To show this, we define, $\mathbf{Q}(t) = \tfrac{\partial}{\partial\gamma} e^{\tilde{\mathbf{\Gamma}}t}$ with the obvious initial condition $\mathbf{Q}(0) = \mathbf{0}$, and write an equation of motion for $\mathbf{Q}(t)$,

$$\tfrac{\partial}{\partial t}\mathbf{Q}(t) = \tfrac{\partial}{\partial t}\tfrac{\partial}{\partial\gamma} e^{\tilde{\mathbf{\Gamma}}t} = \tfrac{\partial}{\partial\gamma}\tfrac{\partial}{\partial t} e^{\tilde{\mathbf{\Gamma}}t} = \tfrac{\partial}{\partial\gamma}\left(\tilde{\mathbf{\Gamma}} e^{\tilde{\mathbf{\Gamma}}t}\right) = \left(\tfrac{\partial}{\partial\gamma}\tilde{\mathbf{\Gamma}}\right) e^{\tilde{\mathbf{\Gamma}}t} + \tilde{\mathbf{\Gamma}}\mathbf{Q}(t). \tag{C16}$$

The Laplace transform of Eq. (C16) is given by,

$$s\overline{\mathbf{Q}}(s) = \left(\tfrac{\partial}{\partial\gamma}\tilde{\mathbf{\Gamma}}\right)\left(s - \tilde{\mathbf{\Gamma}}\right)^{-1} + \tilde{\mathbf{\Gamma}}\overline{\mathbf{Q}}(s),$$

with the solution,

$$\overline{\mathbf{Q}}(s) = \left(s - \tilde{\mathbf{\Gamma}}\right)^{-1}\left(\tfrac{\partial}{\partial\gamma}\tilde{\mathbf{\Gamma}}\right)\left(s - \tilde{\mathbf{\Gamma}}\right)^{-1}. \tag{C17}$$

Noting that, $\partial\vec{\tau}(1)/\partial\gamma = \overline{\mathbf{Q}}(0)\vec{\tilde{P}}_0$ and using Eq. (C17) leads to Eq. (C15). Using the expression for $\partial\vec{\tau}(1)/\partial\gamma$, the derivatives of steady-state probabilities are expressed as,

$$\partial\vec{P}(ss)/\partial\gamma = \tfrac{\partial\vec{\tau}(1)/\partial\gamma}{\vec{U}\vec{\tau}(1)} - \tfrac{\vec{U}\partial\vec{\tau}(1)/\partial\gamma}{\left(\vec{U}\vec{\tau}(1)\right)^2}\vec{\tau}(1)$$

$$= -\frac{\widetilde{\mathbf{\Gamma}}^{-1}\left(\frac{\partial}{\partial\gamma}\widetilde{\mathbf{\Gamma}}\right)\widetilde{\mathbf{\Gamma}}^{-1}\vec{\widetilde{P}}_0}{\vec{U}\widetilde{\mathbf{\Gamma}}^{-1}\vec{\widetilde{P}}_0} + \frac{\vec{U}\widetilde{\mathbf{\Gamma}}^{-1}\left(\frac{\partial}{\partial\gamma}\widetilde{\mathbf{\Gamma}}\right)\widetilde{\mathbf{\Gamma}}^{-1}\vec{\widetilde{P}}_0}{(\vec{U}\widetilde{\mathbf{\Gamma}}^{-1}\vec{\widetilde{P}}_0)^2}\widetilde{\mathbf{\Gamma}}^{-1}\vec{\widetilde{P}}_0 . \quad \text{(C18)}$$

Equation (C18) is easily implemented numerically.

## Tables

| $\phi_{on}(t)$ | | | | $\phi_{off}(t)$ | | | |
|---|---|---|---|---|---|---|---|
| $\{c_{on}\}^{theory}$ | $\{\lambda_{on}\}^{theory}$ | $\{c_{on}\}^{fit}$ | $\{\lambda_{on}\}^{fit}$ | $\{c_{off}\}^{theory}$ | $\{\lambda_{off}\}^{theory}$ | $\{c_{off}\}^{fit}$ | $\{\lambda_{off}\}^{fit}$ |
| 0.2924 | 3.5 | 0.3053 | 3.510 | 0.7280 | 2 | 0.745 | 2.060 |
| -0.0670 | 0.5 | -0.0707 | 0.489 | 0.1112 | 0.2 | 0.1136 | 0.205 |
| 0.0670 | 0.1 | 0.0713 | 0.104 | 0.00160 | 0.02 | 0.0020 | 0.0240 |
| 0.0038 | 0.01 | 0.0036 | 0.0099 | --------- | ------- | ------- | ------- |

**Table 1** The analytical and fit amplitudes and rates in the exponential expansion of $\phi_{on}(t)$ and $\phi_{off}(t)$.

| $\phi_{on+off}(t)$ | | $\phi_{off+on}(t)$ | |
|---|---|---|---|
| $\{\sigma_{on+off}\}^{theory}$ | $\{\sigma_{on+off}\}^{fit}$ | $\{\sigma_{off+on}\}^{theory}$ | $\{\sigma_{off+on}\}^{fit}$ |
| $\begin{pmatrix} -9.70e^{-1} \\ -1.04e^{-2} \\ 2.45e^{-1} \\ 1.21e^{-2} \\ 1.00 \\ -1.89e^{-1} \\ 6.03e^{-3} \end{pmatrix}$ | $\begin{pmatrix} -1.00 \\ -8.38e^{-2} \\ 2.84e^{-1} \\ 1.10e^{-2} \\ 9.98e^{-1} \\ -2.27e^{-1} \\ 2.32e^{-3} \end{pmatrix}$ | $\begin{pmatrix} 9.83e^{-1} \\ -5.85e^{-1} \\ -1.82e^{-3} \\ -1.00 \\ 1.07e^{-1} \\ 4.80e^{-1} \\ 1.73e^{-2} \end{pmatrix}$ | $\begin{pmatrix} 9.98e^{-1} \\ -6.46e^{-1} \\ -4.77e^{-3} \\ -1.00 \\ 1.25e^{-1} \\ 5.31e^{-1} \\ 1.47e^{-2} \end{pmatrix}$ |

**Table 2** The analytical and fit amplitudes in the exponential expansion of $\phi_{on+off}(t)$ and $\phi_{off+on}(t)$.

| $\phi_{(\sqrt{on}+\sqrt{on})^2}(t)$ | | $\phi_{(\sqrt{off}+\sqrt{on})^2}(t)$ | |
|---|---|---|---|
| $\sigma_{on,off}^{theory}$ | $\sigma_{on,off}^{fit}$ | $\sigma_{off,on}^{theory}$ | $\sigma_{off,on}^{fit}$ |
| $1.00$ | $1.00$ | $1.00$ | $1.00$ |
| $-1.54e^{-1}$ | $-1.59e^{-1}$ | $3.35e^{-2}$ | $3.20e^{-2}$ |
| $1.54e^{-1}$ | $1.55e^{-1}$ | $2.20e^{-3}$ | $1.93e^{-3}$ |
| $0.00$ | $1.94e^{-4}$ | $-4.36e^{-2}$ | $-3.49e^{-2}$ |
| $0.00$ | $-9.72e^{-3}$ | $-3.03e^{-2}$ | $-3.16e^{-2}$ |
| $-9.15e^{-3}$ | $-1.21e^{-2}$ | $-9.56e^{-5}$ | $-5.70e^{-4}$ |
| $9.15e^{-3}$ | $1.18e^{-2}$ | $4.36e^{-2}$ | $5.25e^{-2}$ |
| $1.59e^{-3}$ | $1.37e^{-3}$ | $3.03e^{-2}$ | $3.47e^{-2}$ |
| $2.20e^{-3}$ | $3.93e^{-3}$ | $9.56e^{-5}$ | $3.95e^{-5}$ |
| $-3.38e^{-4}$ | $-8.34e^{-5}$ | $1.30e^{-2}$ | $1.38e^{-2}$ |
| $3.38e^{-4}$ | $1.96e^{-4}$ | $4.37e^{-4}$ | $4.13e^{-4}$ |
| $0.00$ | $-2.35e^{-6}$ | $2.86e^{-5}$ | $2.01e^{-5}$ |

**Table 3** The theoretical and fit matrices $\sigma_{x,y}$ s, $x \neq y$. Here, the matrix $\sigma_{x,y}$ is written as a vector of the columns of $\sigma_{x,y}$ put one on top of the other (the entries of the last column of $\sigma_{x,y}$ are the last entries in this vector).

| | |
|---|---|
| $\alpha_{on,j:1}^{theory}$ | $\alpha_{on,j:1}^{fit}$ |
| $\begin{pmatrix} 0 & -8.4e^{-2} & 8.4e^{-2} & 0 \\ 0 & -4.1e^{-2} & 4.1e^{-2} & 0 \end{pmatrix}$ | $\begin{pmatrix} 6.1e^{-2} & -8.8e^{-2} & 8.1e^{-2} & 0 \\ -6.4e^{-2} & -3.7e^{-2} & 4.5e^{-2} & 0 \end{pmatrix}$ |
| $\alpha_{on,j:1}^{theory}$ | $\alpha_{on,j:2}^{fit}$ |
| $\begin{pmatrix} 6.3e^{-1} & 0 & 0 & 0 \\ 0 & 0 & 0 & 8.2e^{-3} \end{pmatrix}$ | $\begin{pmatrix} 5.6e^{-1} & 1.5e^{-2} & -5.1e^{-3} & 1.9e^{-4} \\ 1.2e^{-1} & -6.0e^{-3} & -3.6e^{-3} & 8.6e^{-3} \end{pmatrix}$ |
| $\alpha_{off,j:1}^{theory}$ | $\alpha_{off,j:2}^{fit}$ |
| $\begin{pmatrix} 3.0e^{-1} & 0 & 6.5e^{-4} \\ 1.34 & 0 & 3.0e^{-3} \end{pmatrix}$ | $\begin{pmatrix} 4.42e^{-1} & -1.05e^{-2} & 5.96e^{-4} \\ 1.29 & 3.96e^{-3} & 2.75e^{-3} \end{pmatrix}$ |
| $\alpha_{off,j:1}^{theory}$ | $\alpha_{off,j:2}^{fit}$ |
| $\begin{pmatrix} 0 & 1.6e^{-1} & 0 \\ 0 & 3.6e^{-2} & 0 \end{pmatrix}$ | $\begin{pmatrix} -3.94e^{-2} & 1.71e^{-1} & 1.79e^{-4} \\ 3.15e^{-3} & 2.94e^{-2} & 1.96e^{-4} \end{pmatrix}$ |

**Table 4** The analytical and toolbox-obtained $\{\alpha_{on}, \alpha_{off}\}$.

## Figure Captions

**FIG 1** (color online) A trajectory of an observable that fluctuates between two values, *on* and *off*, as a function of time. Such a trajectory is commonly obtained from single molecule experiments. In this paper, the data is described by a random walk in an *on-off* KS, or its conjugated RD form. Kinetic Monte-Carlo simulations are used to generate the data by a computer. This trajectory was generated by the mapped KS 3. The numerical values for the transitions rates are specified in tables 1 and 4.

**FIG 2** (color online) A set of KSs with only reversible transitions, **A**, **C** and **E**, and the corresponding RD forms, **B**, **D**, and **F**. Although the KSs are very similar, their corresponding RD forms emphasize the differences between these KSs.

**FIG 3** (color online) An irreversible transition KS with three *off* substates and four *on* substates. This is a rank two KS, with a corresponding RD form in Fig. 2D.

**FIG 4** (color online) $\phi_{on}(t)$ (**A**) and $\phi_{off}(t)$ (**B**) for a $10^6$ event trajectory generated by KS 3, on *ln-ln* scale. Shown are the experimentally constructed curves (circles) and the analytical and the fitting curves which practically overlap. The fitting curves are found by our subroutine for exponential fit.

**FIG 5** (color online) **A-** The ordered waiting time trajectory. The $x$ (= *on*, *off*) waiting times are normalized such that the maximal time in the shown interval is unity (this denoted by start symbol). Observed in this figure are *off-off* correlations. **B-** The *off-off* correlation functions for both the actual data (stars) and the randomized data (diamonds), as a function of the distance between events, $m$. Also shown are the error bars as continuous curves. Distinguishing between these correlation functions becomes hard even in its' second argument values, because the error bars overlap. **C-** Moments of successive *on-on* events as a function of the moment's order, $o$. Again, shown are both the results for the actual data (upper points) and the randomized data (lower points), and its corresponding error bars as continuous curves (shown are only the lower error curve for the actual data and the upper error curve for the randomized data). Here, distinguishing between the results from the randomized and the actual data is straightforward even for large values of the power, *o*.

**FIG 6** (color online) **A** – The binned successive WT-PDFs for *on-off* events (lower curve), and the same WT-PDF from the randomized data. The fit for the randomized data is obtained from the amplitudes and rates of the $\phi_x(t)$s. Two fit curves are shown for the actual data (blue and red, online), which overlap almost perfectly with each other and with the experimental curve. These curves are obtained from the direct fitting using our exponential-fit subroutine, and by further translating the found coefficients into the $\sigma_{on+off}$ s. **B** - $\phi_{\left(\sqrt{off}+\sqrt{on}\right)^2}(t)$ for the randomized data (upper curve) and the actual data. The fit for the randomized data is based on the amplitudes and rates of the $\phi_x(t)$s. The fits for the actual data are found from direct fitting, and by further translating the found coefficients into the matrix $\sigma_{on,off}$ s. Here, also, both fit curves overlap with each other and with the experimental curve.

**FIG 7** (color online) Ratios of successive singular values from $\phi_{off,on}(t_1,t_2)$ and its first three cumulative PDFs (**A**), and $\phi_{on,off}(t_1,t_2)$ and its first three cumulative PDFs (**B**). The labels, 0, 1, 2, 3, refer to the order '*n*' of the cumulative WT-PDF, $\phi_{x,y}(t_1,t_2;C_n)$. The actual cutoff is 5.6, see appendix B for details. In both panels, shown are the results from the *third* ratio further, where the values for the second ratios are written on the left (the first ratios are much larger than the second ratios). In both cases, the second order cumulative WT-PDFs give the correct answer of a rank two matrix.

**FIG 8** (color online) The generalized binned successive WT-PDFs $\phi_{\left(\sqrt{x}+\sqrt{y}\right)^2}(t)$ as a function of time, for all four combinations of *x, y=on, off*, on a linear-log scale, obtained from the RD form that was constructed from the data (circles), and the analytical curves (dashed lines). The coincidence is satisfactory in all four panels.

## FIGURES

# Figure 1

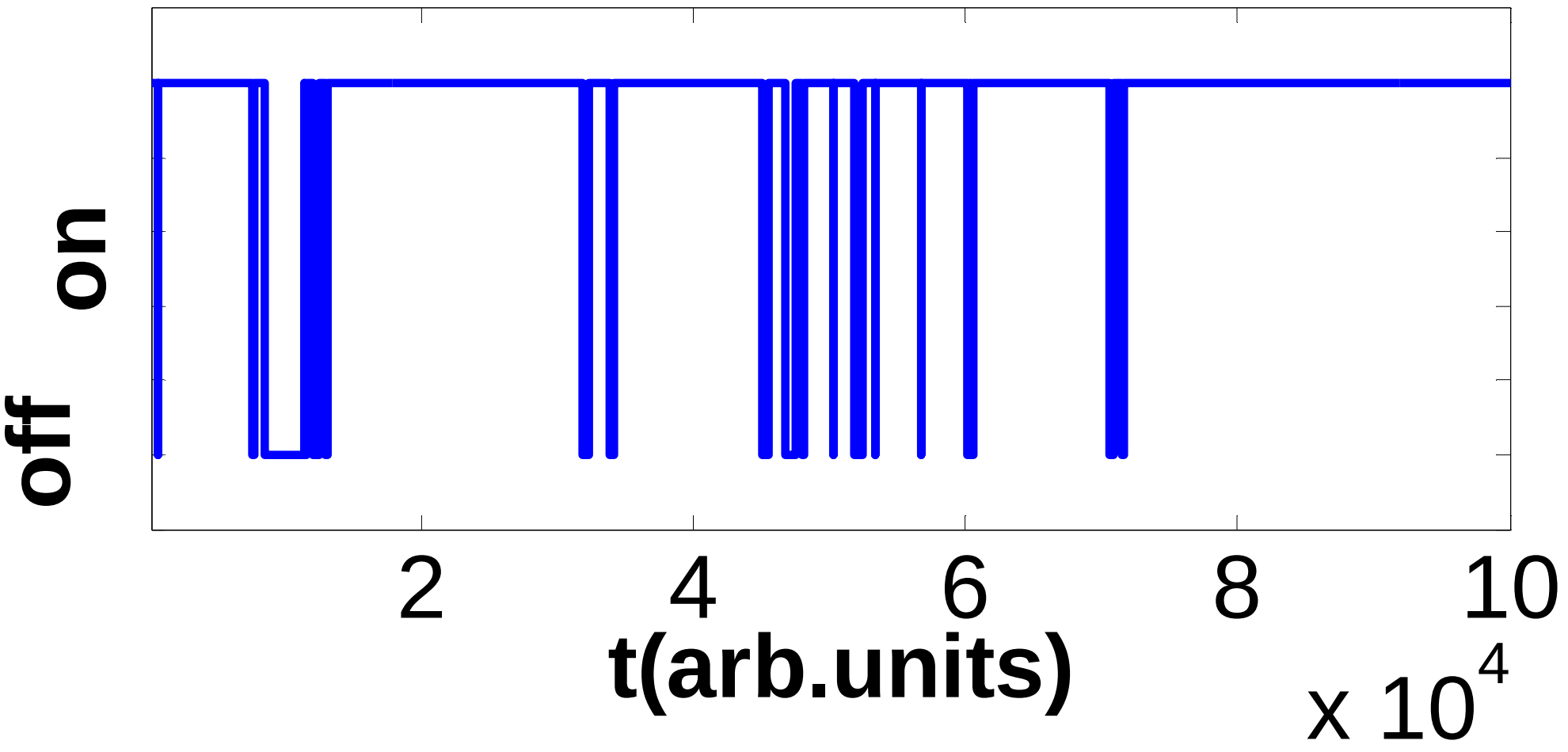

# Figure 2

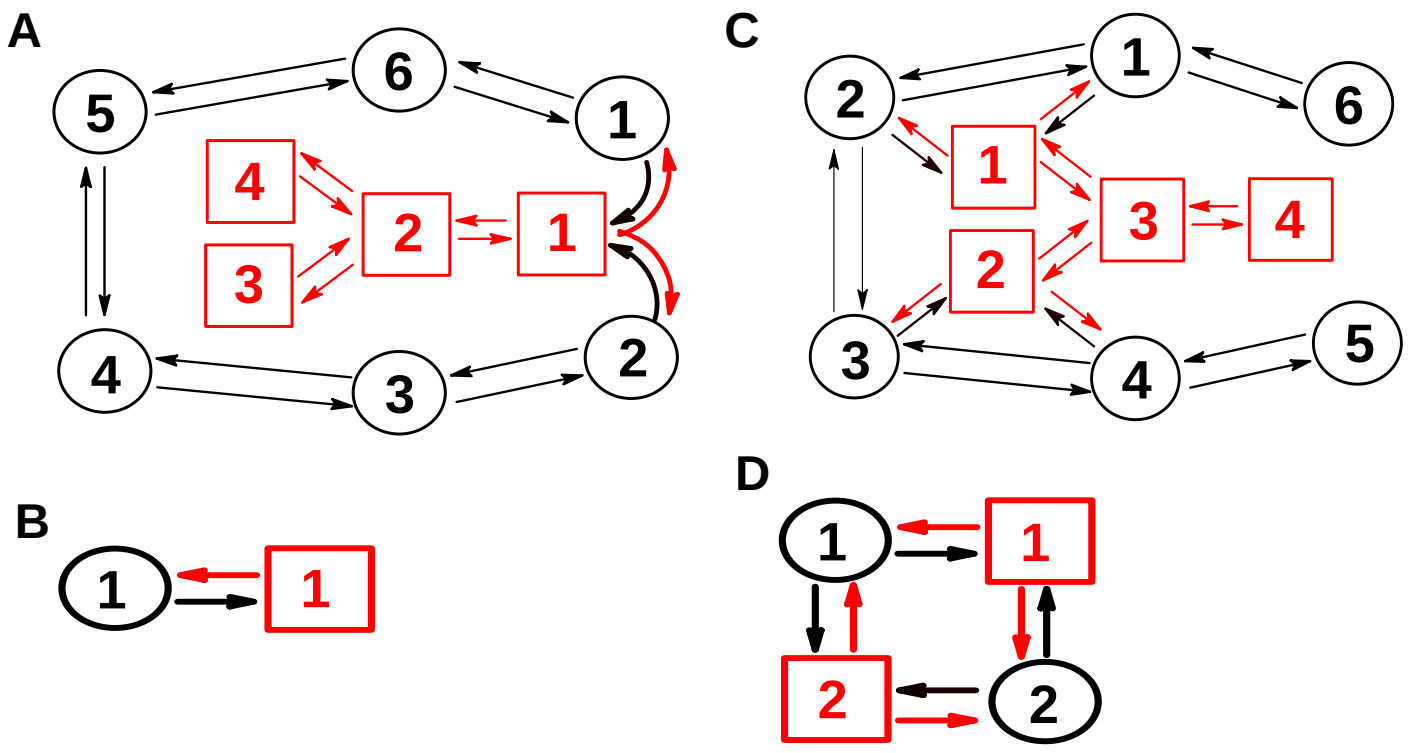

# Figure 3

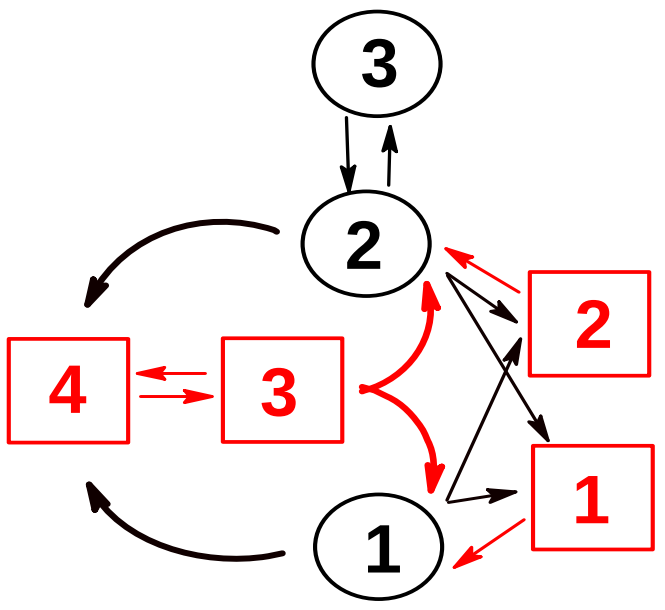

**Figure 4**

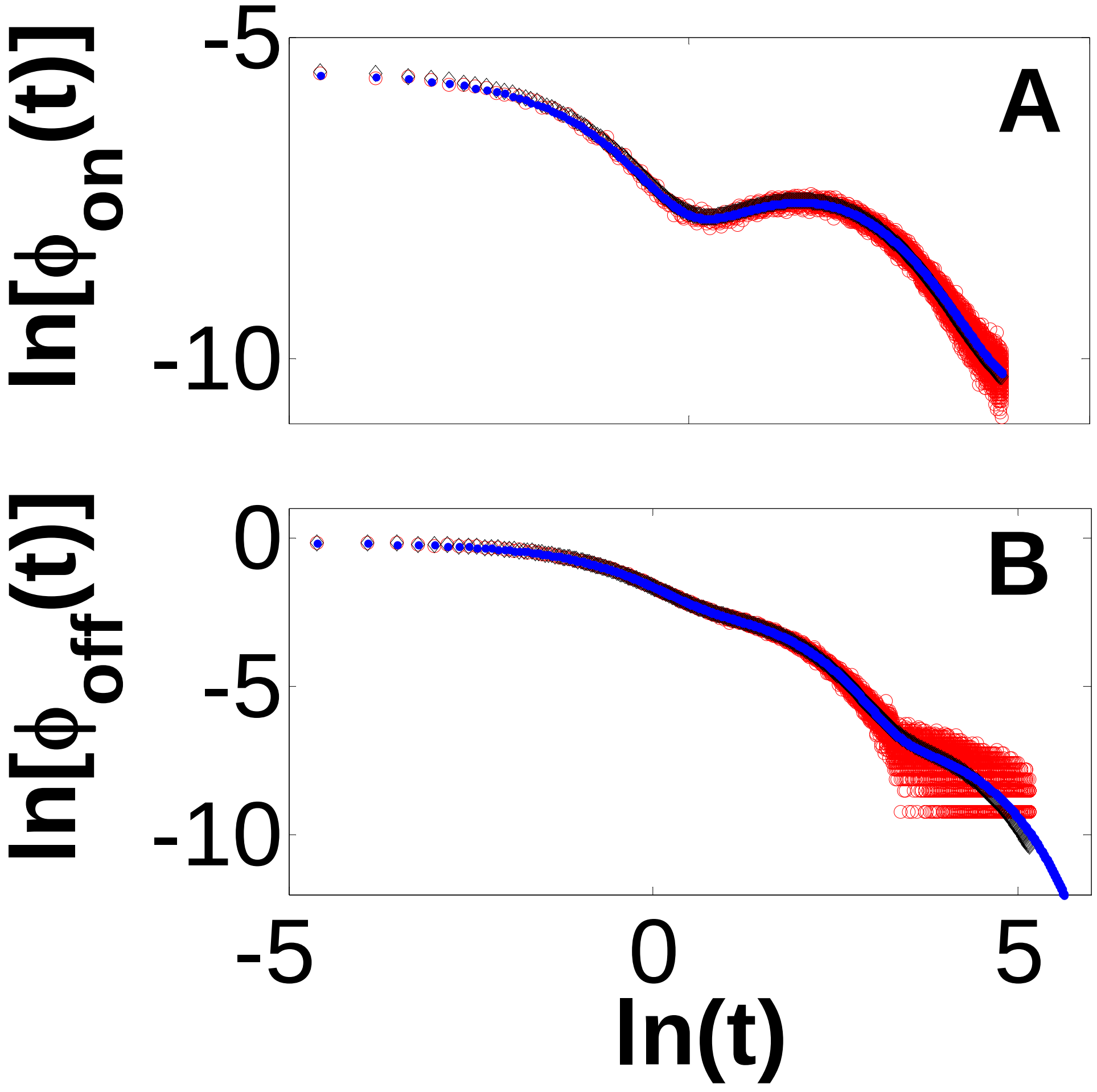

# Figure 5

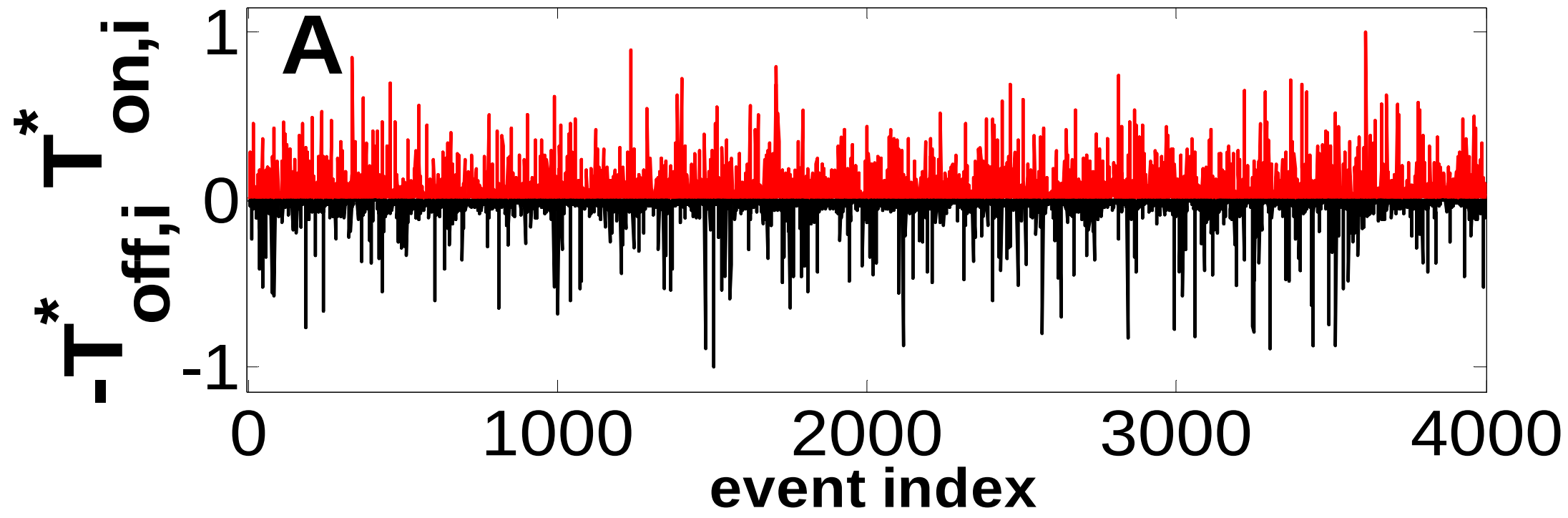

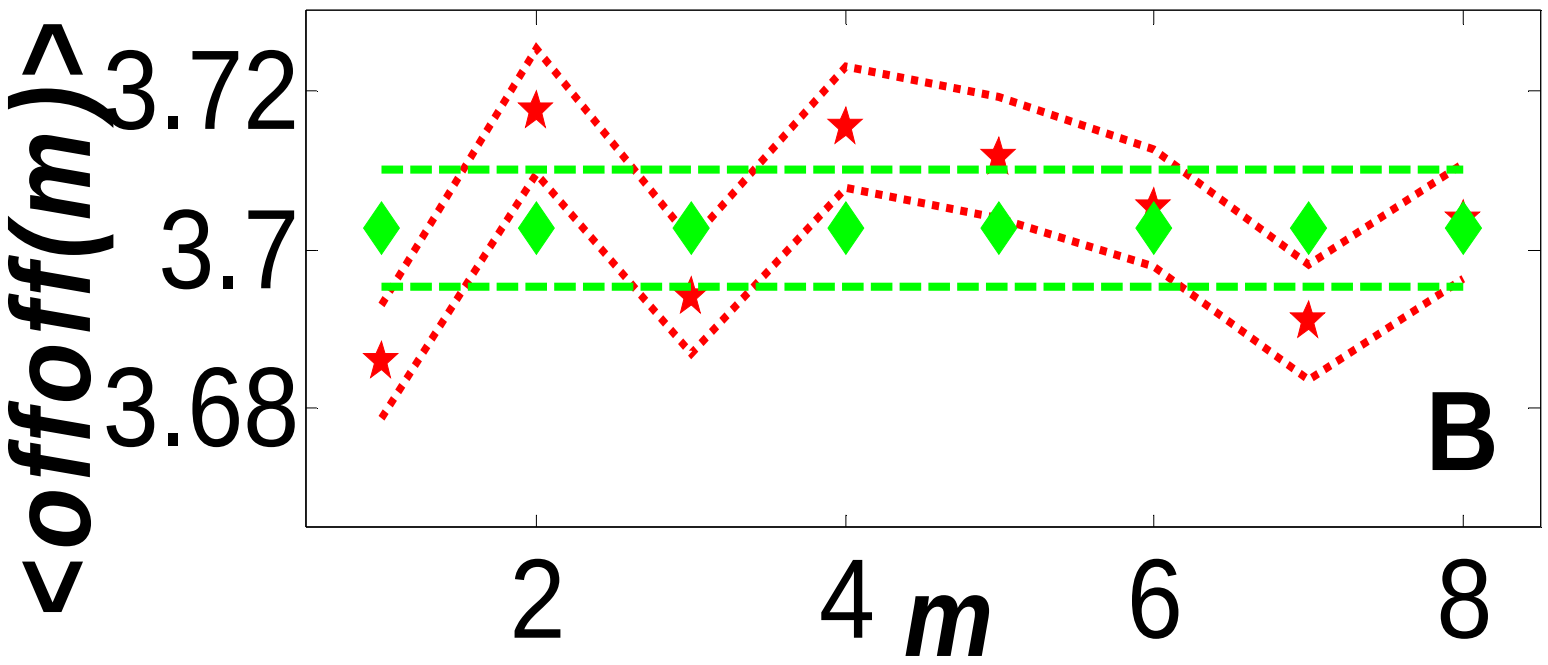

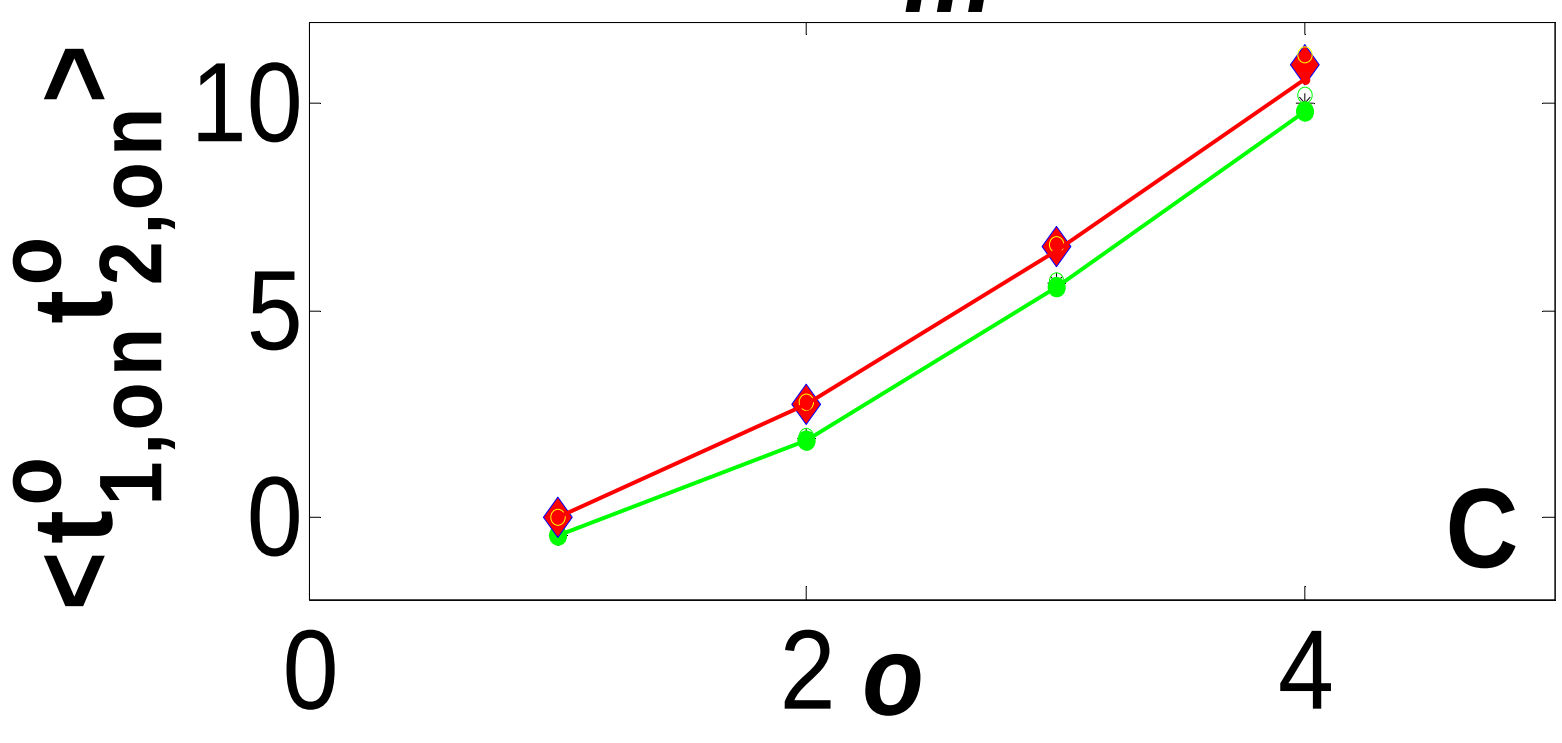

# Figure 6

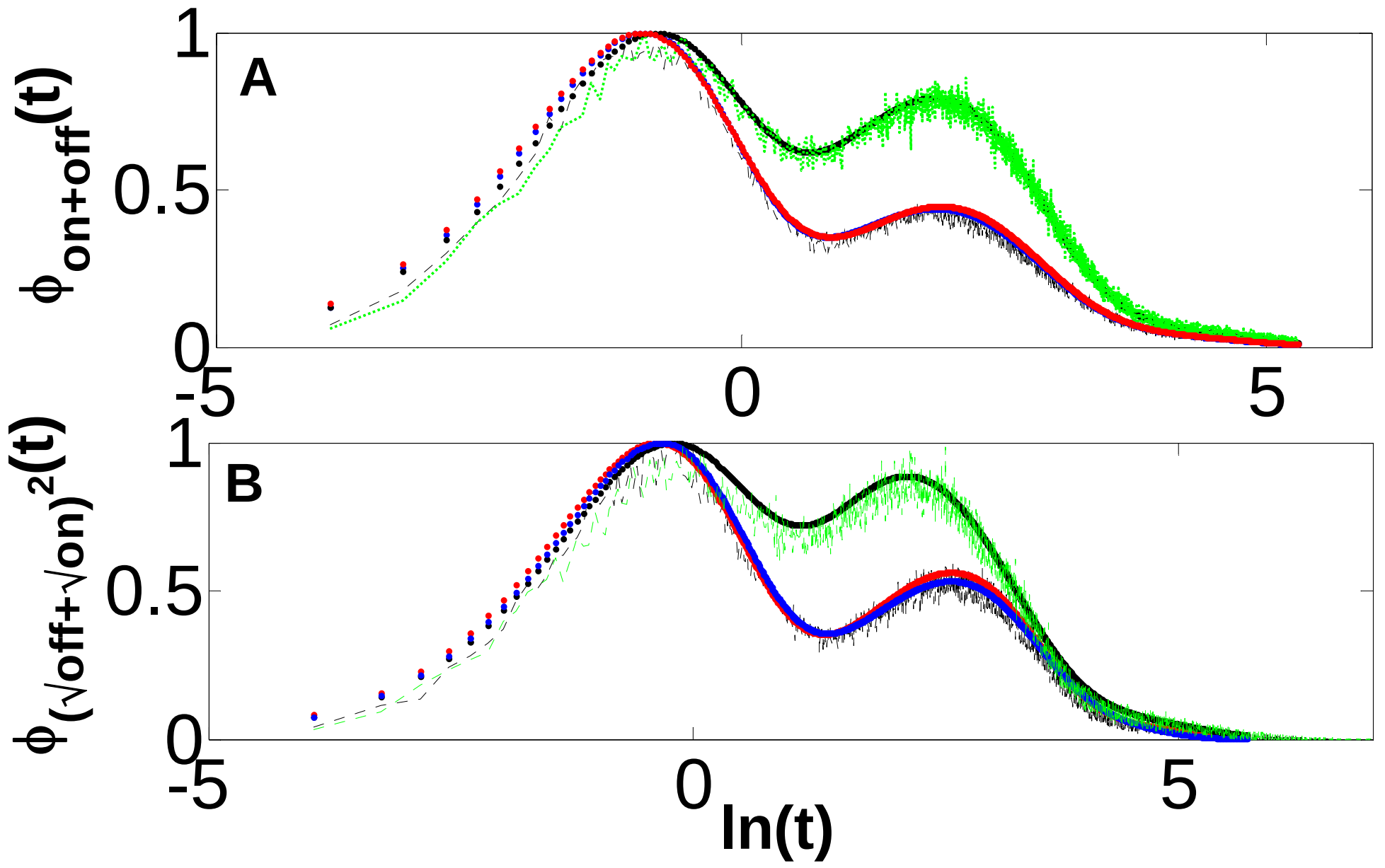

# Figure 7

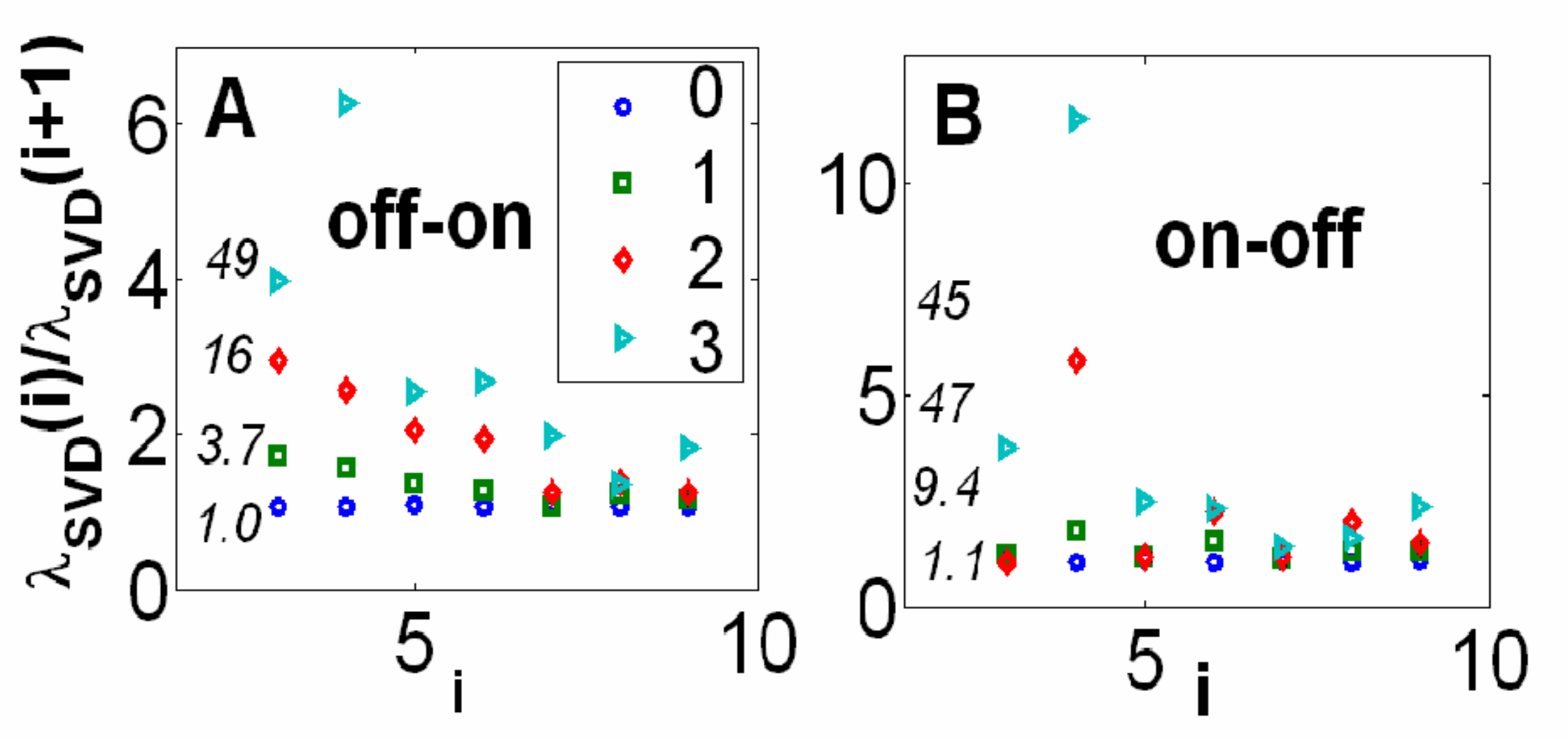

# Figure 8

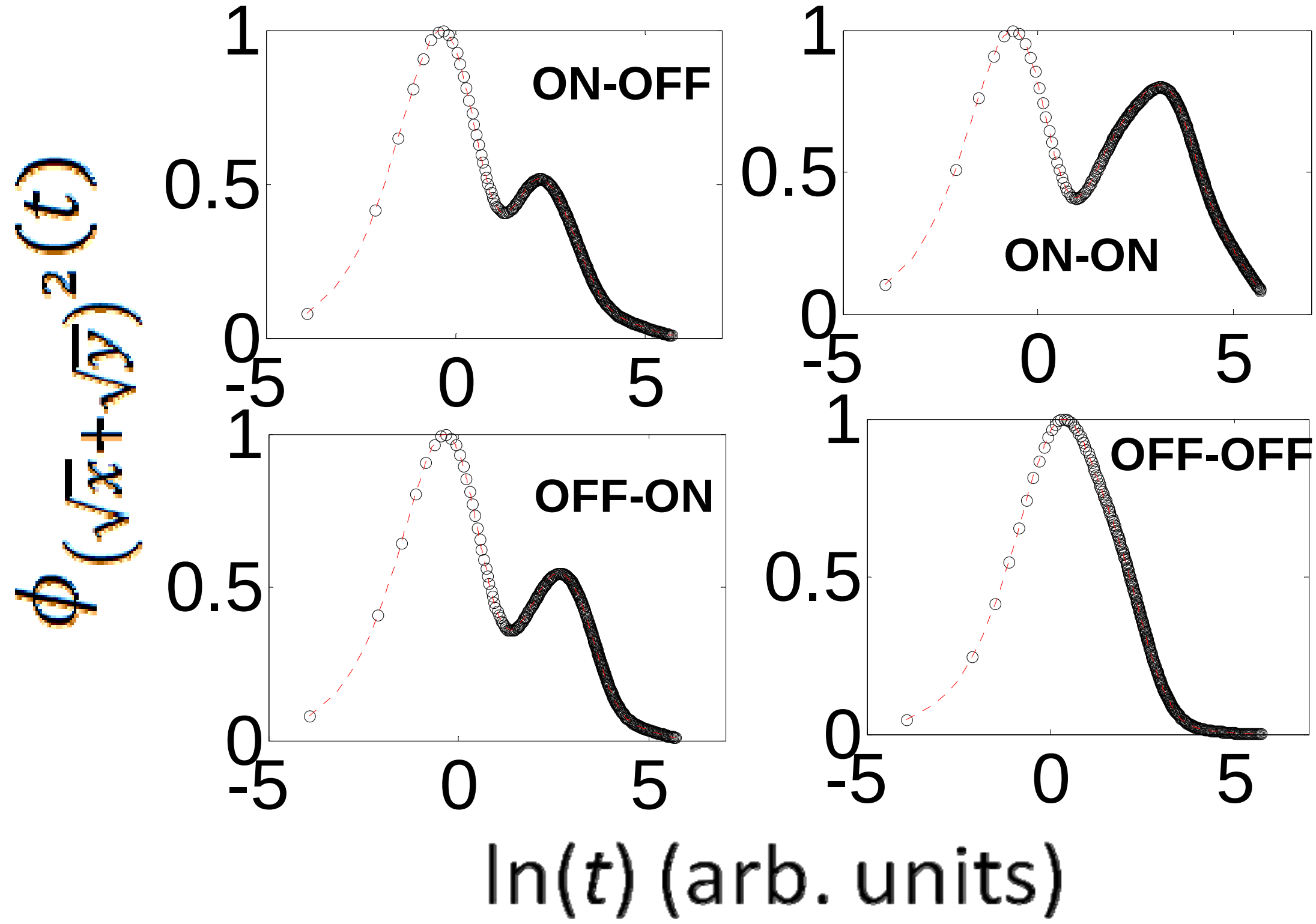